\documentclass[lettersize,journal]{IEEEtran}
\usepackage{amsmath,amsfonts}
\usepackage{array}
\usepackage[caption=false,font=small,labelfont=rm,textfont=rm]{subfig}
\usepackage{textcomp}
\usepackage{stfloats}
\usepackage{url}
\usepackage{verbatim}
\usepackage{graphicx}
\usepackage{cite}
\usepackage{longtable}
\usepackage{xcolor}
\usepackage{amssymb}
\usepackage{algpseudocode}
\usepackage[ruled,linesnumbered]{algorithm2e}

\begin{document}
\bstctlcite{IEEEexample:BSTcontrol}

\title{Multi-UAV Enabled MEC Networks: Optimizing Delay through Intelligent 3D Trajectory Planning and Resource Allocation}

\author{Zhiying Wang, Tianxi Wei, Gang Sun, ~\IEEEmembership{Senior Member, ~IEEE}, Xinyue Liu, \\ Hongfang Yu, ~\IEEEmembership{Senior Member, ~IEEE}, Dusit Niyato, ~\IEEEmembership{Fellow, ~IEEE}
\thanks{This work was supported in part by the National Key Research and Development Program of China under Grant 2019YFB1802800.}
\thanks{Zhiying Wang, Tianxi Wei, Gang Sun, and Hongfang Yu are with the Key Laboratory of Optical Fiber Sensing and Communications (Ministry of Education), University of Electronic Science and Technology of China, Chengdu 611731, China (e-mail: zhiyingwang@std.uestc.edu.cn; 18227273795@163.com; gangsun@uestc.edu.cn; yuhf@uestc.edu.cn).

Xinyue Liu is with the SWJTU-Leeds Joint School, Southwest Jiaotong University, 611756 Chengdu, China (email: cn223xl@leeds.ac.uk).

Dusit Niyato is with the College of Computing and Data Science, Nanyang Technological University, Singapore (e-mail: dniyato@ntu.edu.sg).}}



\maketitle

\begin{abstract}

Mobile Edge Computing (MEC) reduces the computational burden on terminal devices by shortening the distance between these devices and computing nodes. Integrating Unmanned Aerial Vehicles (UAVs) with enhanced MEC networks can leverage the high mobility of UAVs to flexibly adjust network topology, further expanding the applicability of MEC. However, in highly dynamic and complex real-world environments, it is crucial to balance task offloading effectiveness with algorithm performance. This paper investigates a multi-UAV communication network equipped with edge computing nodes to assist terminal users in task computation. 
Our goal is to reduce the task processing delay for users through the joint optimization of discrete computation modes, continuous 3D trajectories, and resource assignment. To address the challenges posed by the mixed action space, we propose a Multi-UAV Edge Computing Resource Scheduling (MUECRS) algorithm, which comprises two key components: 1) trajectory optimization, and 2) computation mode and resource management.
Experimental results demonstrate our method effectively designs the 3D flight trajectories of UAVs, enabling rapid terminal coverage. Furthermore, the proposed algorithm achieves efficient resource deployment and scheduling, outperforming comparative algorithms by at least 16.7$\%$, demonstrating superior adaptability and robustness.
\end{abstract}


\section{Introduction}

\IEEEPARstart{C}{ommunication} technologies exemplified by Beyond 5G (B5G) and 6G are anticipated to drive significant advancements in the Internet of Things (IoT) sector, impacting various aspects including the proliferation of edge devices and their application scenarios \cite{9318751}. However, this rapid escalation in computational demands often contrasts with the limited computational capacity of edge nodes. Mobile Edge Computing (MEC) aims to mitigate this issue by deploying computing resources in proximity to edge devices and delivering computational support tailored to their specific requirements \cite{electronics12173548}. Despite this potential, fixed-location edge servers (ES) are frequently associated with high costs and limited scalability, which restricts their applicability in complex or urgent scenarios. In this context, Unmanned Aerial Vehicles integrated with MEC nodes (UMEC) are emerging as a promising solution due to their inherent flexibility and cost-effectiveness \cite{9452794}. The superior controllability, rapid deployment capabilities, and exceptional mobility of UMEC systems have increasingly captured the attention of researchers, fueling interest across diverse industries in exploring a wide array of contemporary applications \cite{9687317}.

Task offloading typically involves either divisible or atomic task models, resulting in optimization problems that are coupled and non-convex or take the form of mixed-integer nonlinear programming (MINLP). In \cite{Xu2022TCOM}, the authors considered a UMEC network with multiple cellular connections. They investigated a non-convex problem aimed at minimizing weighted energy consumption and obtained local optima by employing block coordinate descent (BCD) and successive convex approximation (SCA). However, these methods often have high computational complexity and are difficult to apply in scenarios with stringent timeliness requirements. In \cite{Lin2023TWC}, the focus shifted to maximizing UAV energy efficiency while ensuring fairness in offloading. To address the MINLP problem with mixed action spaces, the authors used Dueling Deep Q-Network (DQN) and Deep Deterministic Policy Gradient (DDPG) algorithms to manage discrete offloading actions and continuous three-dimensional trajectory decisions, respectively. Despite these advancements, their approach only considered a single agent, making it difficult to scale to large-scale IoT networks. In \cite{10381761}, the authors investigated a binary offloading UMEC system with task prioritization. They used a latent space to represent the actual mixed action space and trained an encoder-decoder architecture using the TD3 algorithm to map between these two spaces. However, single-agent Deep Reinforcement Learning (DRL) methods encounter the curse of dimensionality as the number of drones increases, leading to significantly higher training costs and greater challenges in network convergence. Moreover, the generalizability of this approach requires further investigation, as the relationship between the latent actions and the actual actions is less intuitive compared to the mapping between words and embeddings.


Although significant progress in this field, several critical issues still require in-depth investigation, such as large-scale drone network formation, multi-UAV 3D trajectory planning, and balancing algorithm performance with solution quality in mixed action spaces. To address these challenges, this paper considers the joint binary task offloading problem with multiple drones and combines traditional optimization methods with DRL approaches to strike a balance between efficiency and accuracy. The main contributions of this study are as follows:

\begin{itemize}

\item We introduce a UMEC network model with multi-UAV networking, where UAVs operate in a 3D space without fixed endpoints. This design increases the dimensions and variables to be optimized, offering greater flexibility. The communication links between UAVs not only enhance coverage but also provide computational support. Additionally, we define a delay optimization rate objective that assigns differentiated weights to heterogeneous devices with varying computational capabilities or task sizes, thereby improving the Quality of Experience (QoE) for diverse users.

\item We formulate the problem using a Partially Observable Markov Decision Process (POMDP), which enables us to make decisions based on incomplete information about the environment. Building on this, we propose an intelligent 3D UAV trajectory optimization algorithm that effectively addresses the challenges of planning in dynamic and uncertain environments.

\item We propose a low-complexity algorithm that optimizes offloading decisions and resource allocation strategies, efficiently reducing task processing delays while considering real-time system dynamics and ensuring practical computational complexity.

\item  We analyze the complexity of our proposed algorithms and demonstrate through extensive simulations that our approach outperforms existing algorithms by at least 16.7$\%$ in task delay optimization.

\end{itemize}

The rest of this article is organized as follows. Section II reviews the related work. In Section III, we illustrate the system model and problem formulation. Section IV introduces an alternating algorithm to solve the problem and provides a complexity analysis. Section V provides numerical results to evaluate the performance of our algorithm. Finally, we conclude our work in Section VI.

\section{Related Work}

\subsection{Trajectory design for UAVs}


To provide users with high Quality of Service (QoS) from the perspectives of minimizing latency and energy consumption, the authors in the paper \cite{10279045} address the trajectory planning for a single UAV by the Optimistic Actor-Critic method. In \cite{10620895}, The authors discussed the problem of UAV mobility control and user resource allocation in power-constrained scenarios and divided the problem into three alternating subproblems for resolution. The authors in the \cite{10158431} maximized system energy efficiency by optimizing strategies such as resource scheduling and UAV trajectory while ensuring task completion within the delay constraints. The authors in paper \cite{10333674} minimized network energy consumption by designing UAV trajectory and other parameters while ensuring the computational security of users. The authors in paper \cite{10278822} automated the design of the UAV's 2D flight trajectory to maximize task offloading and provided a reasonable UAV trajectory diagram in the experimental section. The authors in paper \cite{10500359} designed an algorithm based on Spiking Neural Networks (SNN) to compute the UAV trajectory and trained the SNN until convergence. The authors in paper \cite{10284866} jointly minimized the Age of Information (AoI) and energy consumption by designing an offloading strategy and UAV 2D trajectory. The experiments validated the algorithm's convergence and provided the flight trajectory diagram. The authors in paper \cite{10606316} categorized users into two groups based on their activity levels, with both groups' devices able to assist in computation. To address the trajectory design for a single UAV equipped with a wireless charging transmitter, the authors proposed a complex solution method based on Lagrangian duality and successive convex approximation. The authors in paper \cite{10032494} first designed the UAV trajectory and further decided on task offloading and offloading volume based on fairness between UAVs. The starting and ending points of the UAVs' flight were determined, and after reaching the endpoint, the UAVs returned to the starting point for the next round of flight.

\subsection{Resource allocation for tasks}
The authors in paper \cite{9197688} investigated the problem of minimizing the maximum response time in a forest fire monitoring scenario by designing a resource allocation strategy. 
Similarly, the authors in paper \cite{9348573} employed a reinforcement learning algorithm, where each UAV acted as an agent. By optimizing energy allocation and offloading decisions, they minimized user resource consumption while ensuring QoS. In \cite{10592571}, an Artificial Intelligence (AI)-driven framework was introduced, utilizing Deep Learning (DL) to optimize virtualized resource allocation to MEC hosts. This framework has two primary objectives: (i) predicting the UAV density each MEC host must accommodate, and (ii) accurately calibrating the amount of virtual resources required for collision detection applications. In \cite{9345215}, the terminals are vehicles, each with a fixed number of computational tasks to execute within a certain time frame, while UAVs provide computational resources. They first solve UAV deployment and matching using a meta-heuristic method, and then iteratively perform resource allocation for each matching. Furthermore, the authors in paper \cite{9386247} jointly designed resource scheduling strategies, including bit offloading, transmission power, and CPU frequency, to reduce the overall system energy consumption.

\subsection{Minimization of task processing delay}

The authors in paper \cite{9771719} minimized the time required for UAVs to complete tasks by alternately adjusting time scheduling and flight trajectory. The authors in paper \cite{10473780} reduced flight distance by optimizing UAV flight trajectory, direction, and scheduling priority of DAG tasks, thereby significantly lowering task execution delay. The authors in paper \cite{9678008} consider providing computing capabilities for latency-sensitive maritime terminals through a UMEC network. To minimize the total delays, they use two DRL approaches to optimize UAV trajectories and virtual machine configurations, thereby ensuring the efficiency of algorithm execution. The authors in paper \cite{10179258} divided the problem into three layers to minimize energy consumption and delay: 1) task offloading decisions were addressed using game theory, 2) transmission power and UAV trajectory design were solved using the simplistic Geometric Water Filling (GWF) technique and SCA method, and 3) CPU frequency allocation was handled using the gradient descent method. The authors in paper \cite{10431411} considered the problem of minimizing task completion time in a 6G scenario and proposed three solutions, ultimately selecting one method, the Altered Genetic Algorithm (AGA), based on its time complexity. The authors in paper \cite{10216332} minimized the total system delay under energy constraints and proposed a worst-case joint placement and video distribution scheduling problem. The authors in paper \cite{9384267} minimized UAV energy consumption while ensuring user QoE and proposed a multi-stage alternating optimization strategy. Additionally, the authors in paper \cite{9580253} adopted Software-Defined Networking (SDN) technology to enhance user QoS, leveraging its capability to collect network information and make centralized decisions to minimize task processing delay.

Despite progress in UMEC networks, several gaps remain. First, most research uses 2D trajectory planning, limiting UAV mobility and coverage in complex environments, while 3D trajectory optimization is still underexplored. 
Second, most studies focus on optimizing overall performance such as delay, energy, and energy efficiency, which tends to obscure the differences between heterogeneous edge nodes. A more rational approach is to use local execution time as a weight for tasks, allowing for varied optimization weights for edge nodes with different performance levels \cite{Shen2022INFOCOM}.
Third, many solutions struggle to respond quickly to dynamic user demands, lacking the real-time adaptability needed for efficient coverage and minimal delay in fluctuating environments.
To tackle these issues, we propose an effective solution for minimizing the DOR based on the joint optimization of UAV flight paths, binary computation modes, bandwidth, and transmission power to achieve rapid response to user demands.

\section{Problem Description And Modeling}

\subsection{System model}

\begin{figure}[!t]
\centering
\includegraphics[width=3.5in]{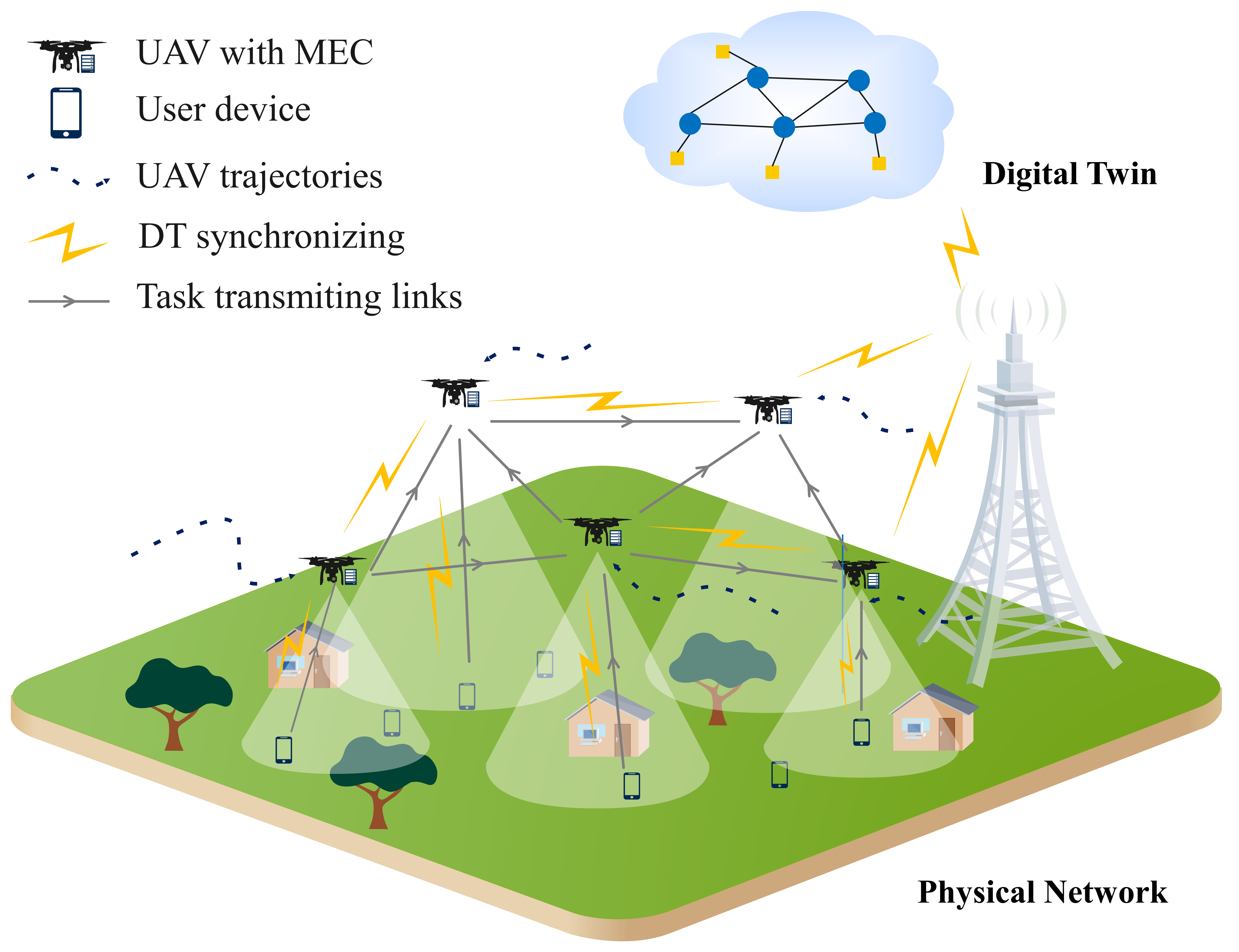}
\caption{System model.}
\label{system_model}
\end{figure}

As shown in Figure \ref{system_model}, we consider a task offloading model in a UMEC network with multi-drone. The network comprises two parts: a physical network and a digital twin network. The system includes $M$ user devices, represented by $\mathcal{M} = \{1,2,...,M\}$, and $N$ UAVs as movable edge servers, represented by $\mathcal{N} = \{1,2,...,N\}$. The position of user $m$ is $\boldsymbol{q}_m(t) = [x_m(t), y_m(t), 0]^T$, while the position of UAV $n$ is $\boldsymbol{r}_n(t) = [x_n(t), y_n(t), z_n(t)]^T$. To achieve long-term optimization of system performance, without loss of generality, we consider a dynamic MEC network over multiple time slots. A continuous period is divided into multiple equal-length time slots, denoted by the set $\mathcal{T} =\left\{0,1,2,\ldots, T\right\}$, with each time slot having a duration of $\delta_t$. In each time slot, each user has an indivisible computation-intensive task $I_m(t) = \{d_m(t), c_m(t)\}$, where the tuple represents the number of bits of the task and the clock cycles required per bit, respectively.

\begin{table}[ht]
\centering
\caption{Explanation of Variables Used in This Paper}
\label{table1}
\begin{tabular}{ll}
\hline
\textbf{Variable} & \textbf{Description} \\ \hline
$a, b$ & Constant parameters related to the environment \\
$B^{G2A}$ & Bandwidth between ground and air \\
$\boldsymbol{B}$ & Set of bandwidth allocations \\
$c$ & Speed of light \\
$C_n^{max}(t)$ & Horizontal radius of UAV $n$'s coverage area \\
$c_m(t)$ & Clock cycles required per bit for $I_m(t)$ \\
$d_m(t)$ & Data size of $I_m(t)$ in bits \\
$D^{min}$ & Safety distance between UAVs \\
$f_{m,n}(t)$ & Processor timeslice assigned to user $m$ by UAV $n$ \\
$f_m(t)$ & Local computation frequency of user $m$ \\
$f_n$ & Maximum CPU frequency of UAV $n$ \\
$f_{r_c}$ & Carrier frequency \\
$\boldsymbol{F}$ & Set of UAV computing resource allocations \\
$\boldsymbol{h}_n(t)$ & Horizontal position of the UAV $n$\\
$I_m(t)$ & Task generated by user $m$ at time $t$ \\
$L_{m,n}(t)$ & Free-space path loss between user $m$ and UAV $n$ \\
$M$ & Number of users \\
$\mathcal{M}$ & Set of users \\
$N$ & Number of UAVs \\
$N_A$ & Environmental noise power in A2A communication \\
$N_G$ & Environmental noise power in G2A communication \\
$\mathcal{N}$ & Set of UAVs \\
$p_m(t)$ & Data transmit power for user $m$ \\
$p_n(t)$ & Data transmit power for UAV $n$ \\
$\boldsymbol{q}_m(t)$ & User $m$' position at timeslot $t$ \\
$\boldsymbol{r}_n(t)$ & UAV $n$' position at timeslot $t$ \\
$r_{m,n}$ & Data transmit rate from user $m$ to UAV $n$ \\
$\boldsymbol{R}$ & Set of UAV trajectories \\
$T$ & Number of total time slots \\
$\mathcal{T}$ & Set of time slots \\
$T^{edge}_{m, n'}(t)$ & Total delay for executing task $I_m(t)$ \\
$T^{exe}_{m, n'}(t)$ & Execution time of task $I_m(t)$ on UAV $n'$ \\
$T^{loc}_{m}(t)$ & Local computation delay of user $m$ \\
$T^{off}_{m, n'}(t)$ & Transmission delay from user $m$ to UAV $n'$ \\
$v^{max}$ & Maximum instantaneous speed of the UAV \\
$\boldsymbol{X}$ & Set of offloading decisions \\
$x_{m, n}(t)$ & Offloading decision variable \\
$X^{max}, Y^{max}$ & Side lengths of the rectangular area \\
$Z^{min}, Z^{max}$ & Minimum and maximum flight altitudes of the UAV \\
$\theta_{m,n}(t)$ & Elevation angle between user $m$ and UAV $n$ \\
$\eta^{LoS}, \eta^{NLoS}$ & Path loss in LoS and NLoS channels \\
\hline
\end{tabular}
\end{table}

\subsection{UAVs movement model}

We restrict the system area to a cubic region, with the coordinates of the origin and the farthest vertex represented as $[0, 0, 0]^T$ and $\left[ X^{max}, Y^{max}, Z^{max} \right]^T$, respectively. All UAVs move horizontally within the region, and to improve coverage, UAVs have a minimum height $Z^{min}$ in the vertical direction. Then we have the following position constraints:

\begin{equation}
\label{x_max}
    0 \leq x_n(t) \leq X^{max}, \forall n\in \mathcal{N}, t \in \mathcal{T}
\end{equation}

\begin{equation}
\label{y_max}
    0 \leq y_n(t) \leq Y^{max}, \forall n\in \mathcal{N}, t \in \mathcal{T}
\end{equation}

\begin{equation}
\label{z_max}
    Z^{min}\leq z_n(t) \leq Z^{max}, \forall n\in \mathcal{N}, t \in \mathcal{T}.
\end{equation}

The movement distance of the UAVs is constrained by the maximum instantaneous speed $v^{max}$ and the minimum collision avoidance distance $D^{min}$, which imposes the following constraints:

\begin{equation}
\label{v_max}
    \Vert \boldsymbol{r}_n(t+1) - \boldsymbol{r}_n(t) \Vert \leq v^{max}\delta_t, \forall n \in \mathcal{N}, t \in \mathcal{T}
\end{equation}
\begin{equation}
\label{collision_constraints}
    \Vert \boldsymbol{r}_n(t) - \boldsymbol{r}_{n^{'}}(t)\Vert \geq D^{min}, \forall n, n^{'} \in \mathcal{N}, n \ne n^{'}.\\[8pt]
\end{equation}

Furthermore, assuming the maximum elevation angle of UAV $n$ is $\phi_n$, the maximum horizontal radius of UAV $n$'s coverage area at time $t$ can be expressed as \cite{9448189,9354996}:

\begin{equation}
    C_n^{max}(t) = z_n(t)\tan(\phi_n), \forall n \in \mathcal{N}, t \in \mathcal{T}.\\[8pt]
\end{equation}

When user $m$ is within the covering region of UAV $n$, the task can be offloaded to UAV $n$. We define the horizontal position of the UAV $n$ as $h_n(t) = [x_n(t), y_n(t), 0]^T$. The constraint is mathematically expressed as:

\begin{equation}
\begin{aligned}
    \label{max_cover}
    x_{m,n}(t)\Vert \boldsymbol{h}_n(t) - \boldsymbol{q}_m(t) \Vert \leq C_n^{max}(t),\\[5pt]
    \forall m\in \mathcal{M}, \forall n\in \mathcal{N}, t \in \mathcal{T}.\\[5pt]
\end{aligned}
\end{equation}

\subsection{Communication model}

As shown in Figure \ref{system_model}, UAVs can receive task offloading requests directly from ground users or from other UAVs in the air. Therefore, this paper considers two communication models: Ground-to-Air (G2A) and Air-to-Air (A2A). Since the results of MEC-processed tasks are typically small, the return link for task result transmission is not considered in this paper.

\subsubsection{G2A}

Due to potential obstructions such as buildings or trees when ground users upload tasks to UAVs, we consider a probabilistic Line-of-Sight (LoS)/Non-Line-of-Sight (NLoS) channel model. The geometric line-of-sight probability between a UAV and a user depends on the environment and the elevation angle. Let $\theta_{m,n}(t)$ denote the elevation angle, then the probabilities of the LoS and NLoS channels between UAV $n$ and user $m$ at time slot $t$ can be expressed as follows:

\begin{align}
    & \mathbb{P}(LoS, \theta_{m,n}(t)) = \frac{1}{1+a\exp{(-b(\theta_{m,n}(t)-a))}} \\
    & \mathbb{P}(NLoS, \theta_{m,n}(t)) = 1 - \mathbb{P}(LoS, \theta_{m,n}(t))
\end{align}\\[5pt]
where $a$ and $b$ are environment-related constant parameters, $\theta_{m,n}(t) = \frac{180}{\pi}\arctan\left(\frac{z_n(t)}{\Vert \boldsymbol{q}_m(t) - \boldsymbol{r}_n(t) \Vert}\right)$. Since a single time slot is typically short, similar to \cite{9687317}, we consider the channel probability condition, and hence the channel probability constant within each time slot. The mean path loss from ground user to air ES is then given by:

\begin{align}
    \overline{PL}_{m,n}(t) & = \mathbb{P}(LoS, \theta_{m,n}(t)) \left( L_{m,n}(t)+\eta^{LoS} \right) \notag \\[8pt] 
    & + \mathbb{P}(NLoS, \theta_{m,n}(t)) \left( L_{m,n}(t)+\eta^{NLoS} \right)
\end{align}\\[5pt]
where $L_{m,n}(t)=20 \lg \left( \Vert \boldsymbol{q}_m(t) - \boldsymbol{r}_n(t)\Vert\right)+20\lg(f_{r_c})-27.56$ represents the free-space path loss, $f_{r_c}$ is the carrier frequency, and $\eta^{LoS}$ and $\eta^{NLoS}$ denote the excess path loss for two kinds of channels.

In this paper, multiple users communicate with UAVs through Orthogonal Frequency Division Multiple Access (OFDMA), while the interference between multiple users within the covering region of a UAV is neglected. Therefore, the expected data transmit rate from user $m$ to UAV $n$ can be given by \cite{10381761}:

\begin{equation}
    r_{m,n} = B^{G2A}_{m,n} \log_2 \left( 1+\frac{p_m(t)}{\overline{PL}_{m,n}(t) N_G } \right)\\[8pt]
\end{equation}
where $p_m(t)$ and $N_G$ indicate the transmit and noise power, respectively, $B^{G2A}_{m,n}$ denotes the uplink bandwidth from user $m$ to UAV $n$. The total uplink transmission bandwidths cannot exceed the maximum ground-to-air transmission bandwidth $B_n^{G2A}$:

\begin{equation}
\label{b_ue}
    \sum_{m \in \mathcal{M}}B^{G2A}_{m,n} \leq B_n^{G2A}, \forall m \in \mathcal{M}, n \in \mathcal{N}, t \in \mathcal{T}\\[8pt]
\end{equation}

\begin{equation}
\label{b_ue_positive}
    B^{G2A}_{m,n} \geq 0, \forall m \in \mathcal{M}, n \in \mathcal{N}, t \in \mathcal{T}.\\[8pt]
\end{equation}

\subsubsection{A2A}
Given that LoS links generally prevail in A2A communication \cite{8873672}, we adopt the free-space path loss model to characterize the communication process from UAV $n$ to $n'$:

\begin{equation}
    PL_{n,n'} = 20 \lg \left( \frac{\Vert \boldsymbol{r}_n(t) - \boldsymbol{r}_{n'}(t) \Vert}{1000} \right) + 20\lg(f_{r_c}) + 32.45 \\[8pt]
\end{equation}
Therefore, the data transfer rate can be given by:

\begin{equation}
    r_{n,n'}(t) = B^{A2A}_{n,n'} \log_2 \left( 1+\frac{p_n(t)10^{-\frac{PL_{n,n'}}{10}}}{N_A}\right)\\[8pt]
\end{equation}
where $p_n(t)$ and $N_A$ indicate the transmit and noise power, respectively, $B^{A2A}_{n,n'}$ denotes the transmit bandwidth from UAV $n$ to $n'$. During task transmission among UAVs, bandwidth is allocated to support parallel transmission. The total transmission bandwidth across UAVs must not exceed the maximum air-to-air bandwidth, $B^{A2A}$, and each bandwidth allocation must be positive, subject to the following constraints:

\begin{equation}
\label{b_uav}
    \sum_{n \in \mathcal{N}}B^{A2A}_{n,n'} \leq B^{A2A}, \quad \forall n \in \mathcal{N}, n' \in \mathcal{N}, t \in \mathcal{T}\\[8pt]
\end{equation}

\begin{equation}
\label{b_uav_positive}
    B^{A2A}_{n,n'} \geq 0, \quad \forall n \in \mathcal{N}, n' \in \mathcal{N}, t \in \mathcal{T}.\\[8pt]
\end{equation}

\subsection{Computation model}

Since task $I_m(t)$ can only be processed in one of two ways: 1) local computation mode, or 2) edge computation mode, we will examine the time costs of both computation paradigms separately. Note that, as there are $N$ UAVs capable of executing computational tasks, task $I_m(t)$ may have $N$ different offloading options in the edge mode. We first define a binary variable $x_{m, n}(t) \in \{ 0, 1 \}, \forall n \in \mathcal{N}^{\dagger} \triangleq \{ 0 \} \cup \mathcal{N}$ to indicate whether task $I_m(t)$ is offloaded to UAV $n$ for execution.

\subsubsection{Local computing mode}

When $x_{m, 0}(t) = 1$, task $I_m(t)$ is executed locally on the user's device. Clearly, local execution does not incur any transmission delay, and the task's response delay depends solely on the local computation delay:

\begin{equation}
T_{m}^{loc}(t)=\frac{d_m(t) \cdot c_m(t)}{f_m(t)}\\[8pt]
\end{equation}
where $f_m(t)$ represents the local computation frequency of user $m$.

\subsubsection{Edge computing mode}

When $x_{m, n^{'}}(t) = 1$, task $I_m(t)$ is offloaded to UAV $n^{'}$ for execution. Given the limited coverage of UAVs, it is necessary to consider whether user $m$ is within the covering region of UAV $n^{'}$. If the user is within the covering region of UAV $n^{'}$, the task offloading process can be completed directly via the G2A link. Otherwise, the task $I_m(t)$ must be offloaded to UAV $n$ via the G2A link and then relayed to UAV $n^{'}$ through the A2A link. Based on the full-duplex communication model, where UAVs can receive tasks from users while simultaneously transmitting tasks to other UAVs, the transmit delay for $I_m(t)$ can be formulated as:

\begin{equation}
    T^{off}_{m, n^{'}}(t) = \max \left\{  \frac{d_m(t)}{r_{m, n}(t)}, \frac{d_m(t)}{r_{n, n^{'}}(t) } \right\}.\\[8pt]
\end{equation}

Since the transmission between UAVs follows a LoS link model with minimal channel interference and the transmission power of UAVs is significantly higher than that of users  \cite{10381761}, i.e., $r_{m, n}(t) \ll r_{n, n^{'}}(t)$, the transmit delay for $I_m(t)$ can be reformulated as:

\begin{equation}
    T^{off}_{m, n^{'}}(t) = \frac{d_m(t)}{r_{m, n}(t)}.\\[8pt]
\end{equation}

Once task $I_m(t)$ is offloaded to UAV $n^{'}$, it can be calculated on the edge node. Let $f_{m,n'}(t)$ represent the processor timeslice allocated to $I_m(t)$, the execution time of $I_m(t)$ on UAV $n^{'}$ can be given by:

\begin{equation}
    T^{exe}_{m, n^{'}}(t) = \frac{d_m(t) \cdot c_m(t)}{f_{m,n'}(t)}.\\[8pt]
\end{equation}

There are also limits to the allocation of computational resources like bandwidth constraints. Specifically, the processor timeslice allocated for tasks executed on UAV $n$ cannot exceed the computational capacity limit of a single UAV:

\begin{equation}
\label{frequence_max}
    \sum _{m\in \mathcal{M}}f_{m, n}(t)\leq f_n, \forall n \in \mathcal{N}, t \in \mathcal{T}\\[3pt]
\end{equation}

\begin{equation}
\label{frequence_positive}
    f_{m, n}(t)\geq 0, \forall n \in \mathcal{N}, t \in \mathcal{T}\\[8pt]
\end{equation}
where $f_n$ represents the maximum CPU frequency of UAV $n$.

In summary, the total delay for task $I_m(t)$ in the edge computation mode is given by:

\begin{equation}
    T^{edge}_{m,n^{'}}(t) = T_{m, n^{'}}^{off}(t) + T^{exe}_{m, n^{'}}(t).\\[8pt]
\end{equation}

\subsection{Problem formulation}

We aim to enhance the long-term QoE of MEC services. The optimization objective is defined as the Delay Optimization Ratio (DOR), which is defined as the ratio of the completion delay of user $m$ in timeslot $t$ to the completion delay without edge computing assistance. The DOR reflects the QoE of users by comparing the task delays with and without MEC assistance and applying unique weights to each user with different computational capabilities and task sizes. The DOR in time slot $t$ can be formulated as:
\begin{equation}
    \begin{aligned}
    O(t) &= \\
    \sum_{m \in \mathcal{M}} &\sum_{n \in \mathcal{N}} \left[1 - \frac{x_{m, n}(t) \cdot T^{edge}_{m, n}(t) + x_{m, 0}(t) \cdot T^{loc}_m(t)}{T^{loc}_m(t)}\right]. 
    \end{aligned}
\end{equation}\\[5pt]

By jointly optimizing the computation mode $\boldsymbol{X} = \{ x_{m, n}(t), \forall m \in \mathcal{M}, n \in \mathcal{N}^{\dagger}, t \in \mathcal{T} \}$, UAV trajectories $\boldsymbol{R} = \{ \boldsymbol{r}_n(t), \forall n \in \mathcal{N}, t \in \mathcal{T} \}$, channel bandwidth allocation $\boldsymbol{B} = \{ B^{G2A}_{m,n}, \forall m \in \mathcal{M}, n \in \mathcal{N}, t \in \mathcal{T} \}$, and computing resource allocation $\boldsymbol{F} = \{ f_{m,n}(t), \forall m \in \mathcal{M}, t \in \mathcal{T} \}$, the DOR maximization problem can be formulated as:

\begin{align}
\text{(P1)} \quad \max_{\boldsymbol{X, R, B, F}} \quad &  \sum_{t \in \mathcal{T}} O(t)  \label{p1} \\[8pt]
\text{s.t.} \quad  (\ref{x_max})-(\ref{collision_constraints}), (\ref{max_cover}&), (\ref{b_ue}), (\ref{b_ue_positive}), (\ref{frequence_max}), (\ref{frequence_positive})  \notag \\[8pt]
 x_{m, n}(t) \in \{ &0, 1 \}, \forall m \in \mathcal{M}, n \in \mathcal{N}^{\dagger}, t \in \mathcal{T} \label{x} \\[8pt]
 \sum^{N+1}_{n=0}x_{m, n}(t) &= 1, \forall m \in \mathcal{M}, n \in \mathcal{N}^{\dagger}, t \in \mathcal{T} \label{sum_x} 
\end{align}
where Constraints (\ref{x_max})-(\ref{collision_constraints}) are UAV position and mobility Constraints, Constraint (\ref{max_cover}) restricts users to offload tasks only to UAVs covering the user, Constraint (\ref{b_ue}) limits the maximum channel bandwidth, Constraint (\ref{b_ue_positive}) ensures the allocated bandwidth is positive, Constraint (\ref{frequence_max}) limits the maximum computing resources of the UAVs, and Constraint (\ref{frequence_positive}) ensures that the computing resources allocated to each UAV are positive. Constraint (\ref{x}) restricts the tasks to a 0-1 offloading strategy, and Constraint (\ref{sum_x}) ensures that each task can only be computed either locally or offloaded to one UAV or base station. Due to the discrete Constraint (\ref{x}) and non-convex objective, P1 is a highly coupled MINLP problem that is challenging to solve directly. Next, we introduce a feasible algorithm that balances execution efficiency and accuracy, based on DRL technology and numerical optimization methods.

\section{Algorithm Design}

\begin{figure}
    \centering
    \includegraphics[width=1\linewidth]{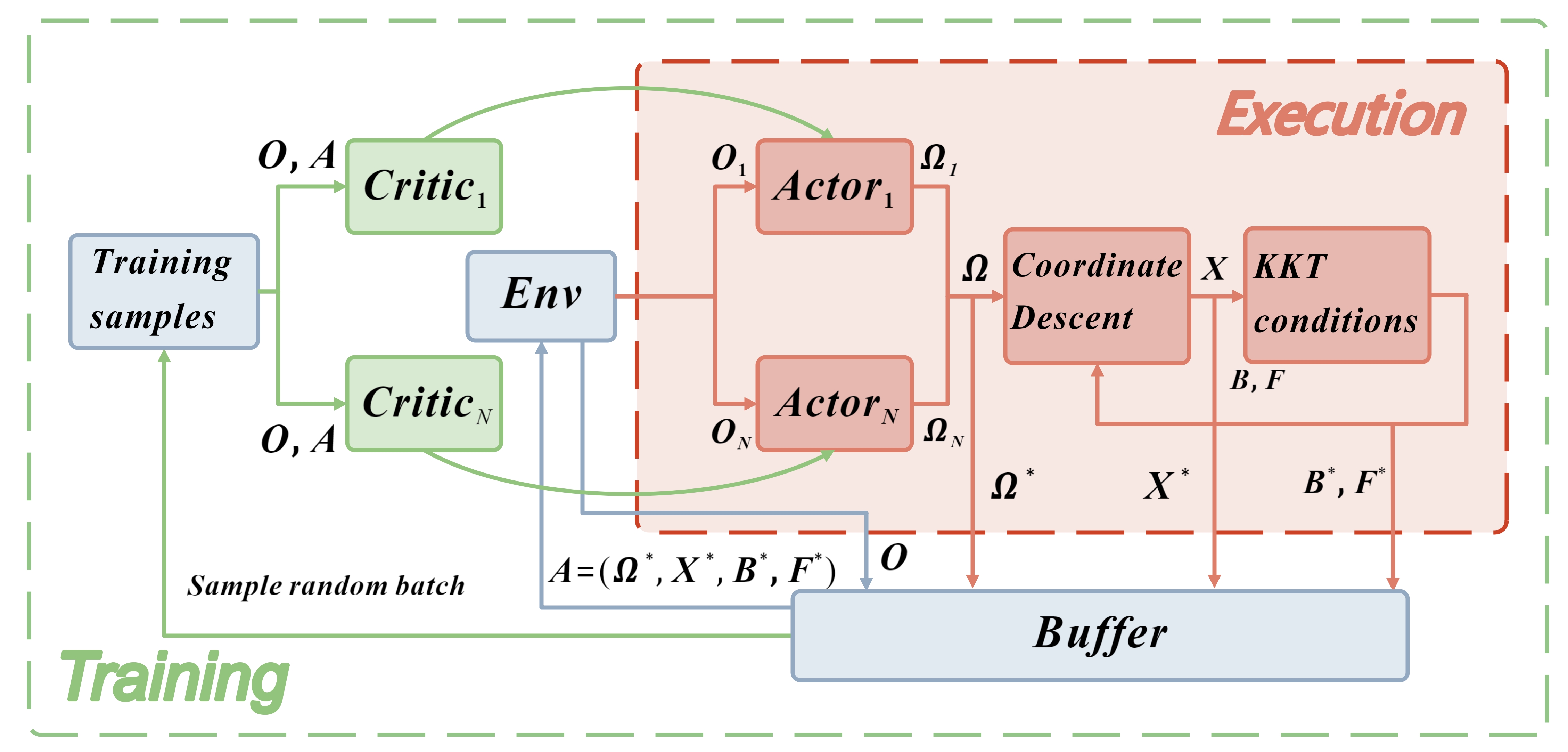}
    \caption{Algorithm framework.}
    \label{Algorithm_model}
\end{figure}

Given the above problem description and modeling, we propose a Multi-UAV assisted Edge Computing Resource Scheduling algorithm (MUECRS). The algorithm model is shown in Figure\ref{Algorithm_model}. The algorithm is primarily divided into two parts: training and execution. We employ the Multi-Agent DDPG (MADDPG) algorithm, where each UAV acts as an intelligent agent. Initially, we initialize two neural networks for each agent: one is the actor network, and the other is the critic network. Upon receiving information from the environment, the agents perform distributed UAV trajectory planning $\boldsymbol{R}$ according to their respective policies. Based on Coordinate Descent (CD) \cite{8334188} and the KKT algorithm, they alternately execute to obtain offloading decisions $\boldsymbol{X}$, bandwidth allocation $\boldsymbol{B}$, and computation resource allocation $\boldsymbol{F}$. All decisions are first returned to the environment for updates, after which the observation and the actions generated by the execution module are stored in the experience buffer. In the training module, we first sample training data from the buffer, update the critic network using temporal difference, and further update the actor network through policy gradients.

\subsection{Trajectory optimization}
We first use a DRL algorithm to compute the UAV trajectory strategy $\boldsymbol{R}$. Problem (P1) can be transformed into:

\begin{align}
\text{(P2)} \quad \max_{\boldsymbol{R}} \quad &  \sum_{t \in \mathcal{T}} O(t)  \label{p2} \\[8pt]
\text{s.t.} \quad  (\ref{x_max})-&(\ref{collision_constraints}). \notag 
\end{align}

\subsubsection{POMDP}

In the UMEC network, the next position of a UAV depends only on its current position and displacement. In problem (P2), we treat each UAV as an agent, and this problem can be represented as a POMDP, which can be described by a tuple: $<\mathcal{U},\mathcal{S}, \mathcal{O}, \mathcal{A}, \mathcal{P}, \mathcal{R},\gamma>$, where $\gamma \in \left[0,1\right)$ is the discount factor. The remaining parameters are explained in detail as follows:
\begin{itemize}
    \item \textbf{Collection of UAV Agents}: 
    Each UAV acts as an agent, $\mathcal{U} = \left\{1, \ldots, U\right\}$, and aims to maximize the DOR by learning the optimal trajectory planning strategy in the current state.
    \item \textbf{State and observation}:
    The state $\mathcal{S}$ of the UAV-assisted edge computing network is derived from the observations $\mathcal{O}$ of each UAV agent. Since each UAV can only observe the state of its own environment, the observation space for UAV $n$ is $o_n=\left\{\boldsymbol{r}_n(t)\right\} \in  \mathcal{O}$, which is the positional information of the UAV itself at the current time. Therefore, the state space is $\mathcal{S} =  \left\{o_1, \ldots, o_N\right\}$.
    \item \textbf{Action}:
To optimize the UAV trajectory, the action of UAV $n$ at time $t$ is defined as $a_n(t) = \left\{\Delta\boldsymbol{r}_n(t)\right\} \in \mathcal{A}$, where $\Delta\boldsymbol{r}_n(t) = \Vert \boldsymbol{r}_n(t+1) - \boldsymbol{r}_n(t) \Vert$ represents the displacement of the UAV at time $t$.
    \item \textbf{Transition Probability}:
$\mathcal{P}(o_n(t+1)|o_n(t),a_n(t))$ denotes the probability of UAV agent $n$ transitioning from state $o_n(t)$ to state $o_n(t+1)$ at time $t$ given action $a_n(t)$.
    \item \textbf{Reward Function}:
Since the objective of the problem (\ref{p1}) is to maximize the DOR by optimizing UAV trajectories, our reward should be positively correlated with the optimization goal. Given that we adopt a cooperative strategy among multiple UAVs, all UAVs share a common reward function:

\begin{equation}
\label{reward}
    \mathcal{R} = 
    \begin{cases}
        O(t), & \text{if constraints are satisfied} \\[8pt]
        O(t) - P_n, & \text{otherwise.}
    \end{cases}
\end{equation}

When the constraints are satisfied, a positive reward is given because the goal is to maximize the DOR. If the constraints are not satisfied, such as in cases where a UAV flies out of bounds or two UAVs collide (high-risk actions), a penalty $P_n$ is imposed into the positive reward.
\end{itemize}

\subsubsection{MADDPG-based training framework}

To address the aforementioned Markov game problem and consider the characteristics of multi-UAV cooperative edge computing, we employ the MADDPG algorithm, which is suitable for the multi-agent-assisted edge computing scenario due to its centralized training and distributed execution features. Each UAV acts as an agent for decision-making, while the base station serves as the training center to evaluate and train the UAVs' decisions, leading to rapid convergence. Since UAV flight trajectories exist in a continuous space, we use the MADDPG algorithm to make continuous action decisions.

Each agent possesses a DDPG model. Let the continuous policy of agent $n$ be $\pi(o_n|\omega_n)$, with the corresponding parameter set $\omega_n$. The actor network of UAV $n$ improves the value function $J$ and updates the corresponding parameters $\omega_n$ using the gradient descent method:

\begin{equation}
\label{actor_update}
    \nabla_{\omega_n}J(\omega_n) = \mathbb{E}[\nabla_{\omega_n}\pi(o_n|\omega_n)\nabla_{a_n}Q(\mathcal{S},\mathcal{A}|\psi_n)|_{a_n=\pi(o_n|\omega_n)}]
\end{equation}
where $Q(\mathcal{S},\mathcal{A}|\psi_n)$ is the centralized actor-critic function, and $\psi_n$ are the corresponding parameters of the critic network.

For the actor network, the parameters $\omega_n$ are updated using the following formula:

\begin{equation}
    \omega_n = \omega_n + \alpha_{\mu}\nabla_{\omega_n}J(\omega_n)\\[8pt]
\end{equation}
where $\alpha_{\mu}$ is the step size.

For the critic network, the loss function to be minimized is:

\begin{equation}
\label{critic_loss}
    L_n = \mathbb{E}[(y_n - Q(\mathcal{S},\mathcal{A}|\psi_n))^2]\\[8pt]
\end{equation}
where the target value $y_n$ is defined as:

\begin{equation}
    y_n = r_n + \gamma Q'(\mathcal{S'},\mathcal{A'}|\theta'_n)|_{a'_n=\pi'(o'_n|\mu'_n)}.\\[8pt]
\end{equation}

The critic network parameters $\psi_n$ are updated as:

\begin{equation}
    \psi_n = \psi_n + \alpha_{\theta}\nabla_{\theta} Q(\mathcal{S},\mathcal{A}|\psi_n).\\[8pt]
\end{equation}

Finally, the soft update for the target networks is performed as:

\begin{equation}
\label{soft_update}
\left\{
\begin{aligned}
\psi_n' &\leftarrow \tau \psi_n + (1 - \tau) \psi_n'  \\[8pt]
\omega_n' &\leftarrow \tau \omega_n + (1 - \tau) \omega_n' .
\end{aligned}
\right.
\end{equation}

\subsection{Offloading decision and resource allocation}

After computing the UAV displacements using the MADDPG algorithm, the optimization problem (P1) can be transformed into:

\begin{align}
\text{(P3)} \quad \max_{\boldsymbol{X, B, F}} \quad &  \sum_{t \in \mathcal{T}} O(t)  \label{p3} \\[8pt]
\text{s.t.} \quad & (\ref{b_ue}),(\ref{b_ue_positive}), (\ref{frequence_max}), (\ref{frequence_positive}), (\ref{x}), (\ref{sum_x}).\notag 
\end{align}

However, since the optimization variables of the problem (P3) include both discrete offloading decisions $\boldsymbol{X}$ and continuous bandwidth allocation $\boldsymbol{B}$ and computing resource allocation $\boldsymbol{F}$, this problem is a highly coupled non-convex problem. Considering the complexity of the optimization problem and the low-delay requirements of UAV scenarios, we design a low-complexity solution based on CD and KKT conditions.

In the process of alternately finding the optimal decision for each user, we first fix the offloading decisions $\boldsymbol{X}$ and simplify problem (P3) into:

\begin{align}
\text{(P4)} \quad \max_{\boldsymbol{B, F}} \quad &  \sum_{t \in \mathcal{T}} O(t)  \label{p4} \\[8pt]
\text{s.t.} \quad & (\ref{b_ue}), (\ref{b_ue_positive}), (\ref{frequence_max}), (\ref{frequence_positive}).\notag 
\end{align}

Problem (P4) is a convex problem, so we solve the bandwidth allocation $\boldsymbol{B}$ and computing resource allocation $\boldsymbol{F}$ using KKT conditions, which have low algorithmic complexity. By fixing the UAV displacements and offloading decisions and treating them as constants, Equation (\ref{p4}) can be simplified and expanded into a form solvable by KKT:

\begin{align}
\min_{\boldsymbol{B, F}} \quad &  \sum_{t \in \mathcal{T}}\sum_{m \in \mathcal{M}} \sum_{n \in \mathcal{N}}x_{m, n}(t) \cdot \frac{T^{exe}_{m, n^{'}}(t)}{T^{loc}_m(t)}  \\[8pt]
\text{s.t.} \quad  \sum_{\mathcal{M}}&B^{G2A}_{m,n}-B_n^{G2A} \leq 0, \forall m \in \mathcal{M}, n \in \mathcal{N}, t \in \mathcal{T}\label{g1}\\[8pt]
-B&^{G2A}_{m,n} \leq 0, \forall m \in \mathcal{M}, n \in \mathcal{N}, t \in \mathcal{T}\label{g2}\\[8pt]
\sum& _{i=1}^{M}f_{m, n}(t)-f_n\leq 0, \forall n \in \mathcal{N}, t \in \mathcal{T}\label{g3}\\[8pt]
-f&_{m, n}(t)\leq 0, \forall n \in \mathcal{N}, t \in \mathcal{T}.\label{g4}
\end{align}

At this point, we introduce the Lagrange function:

\begin{equation}
\begin{aligned}
   &\mathcal{L}(\mathbf{B,F}, \boldsymbol{\lambda_1, \lambda_2, \lambda_3, \lambda_4})\\[8pt]
    & = \sum_{t \in \mathcal{T}} \sum_{m \in \mathcal{M}} \sum_{n \in \mathcal{N}} x_{m, n}(t) \cdot 
    \frac{T^{exe}_{m, n^{'}}(t)}{T^{loc}_m(t)} \\[8pt]
    & + \boldsymbol{\lambda_1} \left( \sum_{m \in \mathcal{M}} B^{G2A}_{m,n} - B_n^{G2A} \right) 
    + \boldsymbol{\lambda_2} \left( -B^{G2A}_{m,n} \right) \\[8pt]
    & + \boldsymbol{\lambda_3} \left( \sum_{i=1}^{M} f_{m, n}(t) - f_n \right) 
    + \boldsymbol{\lambda_4} \left( -f_{m, n}(t) \right),\\[8pt]
    &\quad\forall m \in \mathcal{M}, n \in \mathcal{N}, t \in \mathcal{T}.
\end{aligned}
\end{equation}

The KKT conditions for problem (P4) are obtained by considering the Lagrange function's stationarity, primal feasibility, dual feasibility, and complementary slackness conditions:

\[
\begin{cases}
\frac{\partial \mathcal{L}}{\partial {B^{G2A}_{m,n}}} = 0, \quad \forall m \in \mathcal{M}, n \in \mathcal{N}\\[8pt]
\frac{\partial \mathcal{L}}{\partial {f_{m, n}(t)}} = 0, \quad \forall m \in \mathcal{M}, n \in \mathcal{N}\\[8pt]
\boldsymbol{\lambda_1} \left( \sum_{m \in \mathcal{M}} B^{G2A}_{m,n} - B_n^{G2A} \right)  = 0, \quad \forall m \in \mathcal{M}, n \in \mathcal{N}\\[8pt]
\boldsymbol{\lambda_2} \left( -B^{G2A}_{m,n} \right) = 0, \quad \forall m \in \mathcal{M}, n \in \mathcal{N}\\[8pt]
\boldsymbol{\lambda_3} \left( \sum_{i=1}^{M} f_{m, n}(t) - f_n \right) = 0, \quad \forall m \in \mathcal{M}, n \in \mathcal{N}\\[8pt]
\boldsymbol{\lambda_4} \left( -f_{m, n}(t) \right) = 0, \quad \forall m \in \mathcal{M}, n \in \mathcal{N}\\[8pt]
\boldsymbol{\lambda_i} \geq 0, \quad i = 1,2,3,4\\[8pt]
(\ref{g1})-(\ref{g4}).
\end{cases}
\]

From the aforementioned KKT conditions, the optimal bandwidth and computing resource allocation can be determined as follows:

\begin{equation}
\label{bandwidth_best}
B_{m,n}^{G2A*} = B_n^{G2A} \cdot \frac{\sqrt{\displaystyle \frac{f_m(t)}{c_m(t)\log_2\left(1+\frac{p_m(t)}{P_{L_{m,n}}(t)N_G}\right)}}}{\displaystyle \sum_{m \in \mathcal{M}_n} \sqrt{\displaystyle \frac{f_m(t)}{c_m(t)\log_2\left(1+\frac{p_m(t)}{P_{L_{m,n}}(t)N_G}\right)}}}\\[8pt]
\end{equation}

\begin{equation}
\label{frequency_best}
    {f_{m, n}(t)}^* = f_n \cdot \frac{ \sqrt{f_m(t)}}{\displaystyle \sum_{m \in \mathcal{M}_n} \sqrt{f_m(t)}}\\[8pt]
\end{equation}
here, ${\mathcal{M}_n}$ represents the set of users that offload tasks to UAV $n$.

Additionally, since the offloading decision follows a 0-1 strategy, using an exhaustive search to solve the problem would result in an algorithmic complexity of $O(2^{MN})$. As the number of users $M$ and the number of UAVs $N$ increase, the computational complexity grows exponentially, making it infeasible to exhaustively search through all possible offloading decisions for large-scale problems. Therefore, we adopt a low-complexity one-directional search method. We define the offloading decision at the $l$-th iteration as:

\begin{equation}
\label{iteration_X}
    \boldsymbol{X}^l = \left\{\boldsymbol{X}^l_1(t), \boldsymbol{X}^l_2(t),\ldots,\boldsymbol{X}^l_m(t),\ldots, \boldsymbol{X}^l_M(t)\right\}\\[8pt]
\end{equation}
here, $\boldsymbol{X}^l_m(t) = \left\{x^l_{m,0}(t), x^l_{m,1}(t),x^l_{m,2}(t),\ldots, x^l_{m,N}(t)\right\}$ represents the offloading decision of user $m$ in the $l$-th iteration. When $l=0$, users initially choose local offloading, i.e., $x^0_{m,0}=1$.

Initially, we set all users to choose the local computation strategy, meaning for all users $m \in \mathcal{M}$, $x_{m,0}^0(t)=1$. In each iteration, we fix the offloading decisions of all other users and change only one user's offloading decision. In the $l$-th iteration, for each user $m$, we sequentially traverse its offloading decisions as:

\begin{equation}
\label{x_update}
    \boldsymbol{X}^l_{m,n}(t) = \left\{0,0,\ldots, 1,\ldots,0\right\}, n=0,1,\ldots, N\\[8pt]\\[8pt]
\end{equation}
where the $n$-th position is selected for offloading, and the other positions are not. Let $O^l_{m,n}(t)$ denote the DOR when user $m$ offloads its task to UAV $n$ in the $l$-th iteration. We select the optimal offloading decision for user $m$ and update it as follows:

\begin{equation}
\label{x_update_best}
    \boldsymbol{X}^{*l}_{m,n}(t) = \arg\max \left\{O^l_{m,1}(t),O^l_{m,2}(t),\ldots, O^l_{m,N}(t)\right\}\\[8pt]
\end{equation}

\begin{equation}
    \begin{aligned}
        \boldsymbol{X}^l = \left\{\boldsymbol{X}^{*{l}}_{1,n}(t),\ldots, \boldsymbol{X}^{*{l}}_{m-1,n}(t),\boldsymbol{X}^{*l}_{m,n}(t),\right.\\[8pt]
        \left.\boldsymbol{X}^{*{l-1}}_{m+1,n}(t),\ldots, \boldsymbol{X}^{*{l-1}}_{M,n}(t)\right\}.
    \end{aligned}
\end{equation}\\[5pt]

After updating the offloading decision of the last user $\boldsymbol{X}^{*{l}}_{M,n}(t)$, we complete one iteration. The stopping criterion for the iterations is $\boldsymbol{X}^l = \boldsymbol{X}^{l-1}$, meaning that if the offloading decision updated in the current iteration is the same as that in the previous iteration, we stop the iterations and output the optimal offloading decision $\boldsymbol{X^*} = \boldsymbol{X}^l$.

Thus, we propose a low-complexity solution based on CD and KKT conditions to optimize task offloading decisions and network resource allocation. This method alternately optimizes offloading decisions and resource allocation during each user's search for the optimal decision and finds a near-optimal solution within a limited number of iterations, resulting in a high-performance solution that meets the low-delay requirements of UAV scenarios. The details are summarized in \textbf{Algorithm 1}.
\begin{algorithm}[t]
\caption{Offloading Decision and Resource Allocation}
\KwIn{UAV trajectories $\boldsymbol{R}$, user tasks, and system parameters.}
Initialize offloading decisions $\boldsymbol{X}^0_m(t)$ according to Equation (\ref{iteration_X})\;
\While{CD not converged}{
    \For{each user $m$}{
        \For{each UAV $n$}{
            Update offloading decision $X^l_{m,n}(t)$ according to Equation (\ref{x_update})\;
            Update optimal bandwidth $\boldsymbol{B}^{*l}$ and computing resources $\boldsymbol{F}^{*l}$ using KKT conditions in Equations (\ref{bandwidth_best}) and (\ref{frequency_best})\;
            Calculate delay optimization $O^l_{m,n}(t)$\;
        }
        Update offloading decision $X^{*l}_{m,n}(t)$ with maximum $O^l_{m,n}(t)$ according to Equation (\ref{x_update_best})\;
    }
}
\KwOut{Optimal offloading decisions $\boldsymbol{X}^*$, bandwidth allocation $\boldsymbol{B}^*$, and computing resources $\boldsymbol{F}^*$.}
\end{algorithm}
\begin{algorithm}[t]
\caption{Multi-UAV assisted Edge Computing Resource Scheduling}
\KwIn{Initial UAV positions $\boldsymbol{r}_n(t)$, actor-critic networks for each UAV, replay buffer $B$.}
\For{episode $= 1$ to $E$}{
    \For{each time slot $t$}{
        \For{each UAV $n$}{
            Observe state $o_n(t)$\;
            Select action $a_n(t) = \pi(o_n(t)|\omega_n)$ using policy $\pi(\cdot|\omega_n)$\;
            Execute action $a_n(t)$ and update UAV position $\boldsymbol{r}_n(t+1)$\;
            Based on $\boldsymbol{r}_n(t+1)$, get the optimal offloading decisions $\boldsymbol{X}^*$, bandwidth allocation $\boldsymbol{B}^*$, and computing resources $\boldsymbol{F}^*$ from \textbf{Algorithm 1}\;
            Calculate reward $R_n(t)$ based on Equation (\ref{reward})\;
            Observe new state $o_n(t+1)$ and store transition $(o_n(t), a_n(t), R_n(t), o_n(t+1))$ in replay buffer $B$\;
        }
        \If{replay buffer $B$ is ready for training}{
            \For{each UAV $n$}{
                Sample a random mini-batch of transitions from $B$\;
                Minimize the loss function (\ref{critic_loss}) to update the critic network\;
                Apply the sampled policy gradient (\ref{actor_update}) to update the actor network\;
                Perform a soft update on the target networks following Equation (\ref{soft_update})\;
            }
        }
    }
}
\KwOut{Optimal UAV trajectories $\boldsymbol{R}^*$, offloading decisions $\boldsymbol{X}^*$, and resource allocations $\boldsymbol{B}^*$, $\boldsymbol{F}^*$.}
\end{algorithm}
\subsection{Complexity analysis}
The procedure for the MUECRS is outlined in \textbf{Algorithm 2}. Additionally, we will conduct a complexity analysis of the algorithm.

\subsubsection{MADDPG algorithm complexity analysis}
The MADDPG algorithm is used to compute the flight trajectories of UAVs. Its computational complexity and space complexity can be estimated based on the neural network architecture. For each agent, there are four neural networks: two actor networks and critic networks.

The complexity of a single forward and backward propagation through the network is mainly determined by the matrix multiplication in each layer. The input state space dimension for the actor network is 3, the output action space dimension is 3, and there are two hidden layers in the fully connected neural network, each with $e_a$ neurons. The computational complexity of the actor network is $\mathcal{O}_{actor} = \mathcal{O}(3\times e_a + 2\times {e_a}^2+e_a\times3)$.
The input to the critic network includes both the state space and the action space, with a dimension of 6, and the output value dimension is 1. There are two hidden layers in the fully connected neural network, each with $e_c$ neurons. The computational complexity of the critic network is $\mathcal{O}_{critic} = \mathcal{O}(6\times e_c + 2\times {e_c}^2+e_c\times1)$.

The space complexity is primarily determined by the weight matrices and bias vectors of the network.
The space complexity of the actor network is $\mathcal{O}_{actor} = \mathcal{O}(3\times e_a + e_a + 2\times({e_a}^2+e_a)+e_a\times3 + 3)$.
The space complexity of the critic network is $\mathcal{O}_{critic} = \mathcal{O}(6\times e_c + e_c + 2\times({e_c}^2+e_c)+e_c\times1 + 1)$.
The experience replay buffer is used to store $B$ experience samples. Since each sample includes the state, action, reward, and next state, the total space complexity of the experience replay buffer is $\mathcal{O}(8B)$.

\subsubsection{CD and KKT algorithm complexity analysis}
Since we sequentially traverse all possible offloading decisions for a single user and select the optimal decision, the algorithm complexity for each user is $\mathcal{O}(N)$, and the algorithm complexity for one iteration is $\mathcal{O}(MN)$. Let $L$ be the number of CD iterations, then the complexity of the CD algorithm is $\mathcal{O}(LMN)$. Additionally, since we compute the optimal bandwidth allocation and resource allocation using the KKT conditions, the algorithm complexity for the KKT algorithm is $\mathcal{O}(1)$. Therefore, the total complexity is $\mathcal{O}(LMN)$.

\section{Simulation Results And Analysis}

In this section, we conduct a comparative analysis of the algorithm proposed in this paper against existing state-of-the-art algorithms. First, we provide a detailed introduction to the experimental parameter settings, followed by the presentation of the experimental results and an in-depth analysis of those results.

\subsection{Parameter settings}
\begin{figure}
    \centering
    \includegraphics[width=1\linewidth]{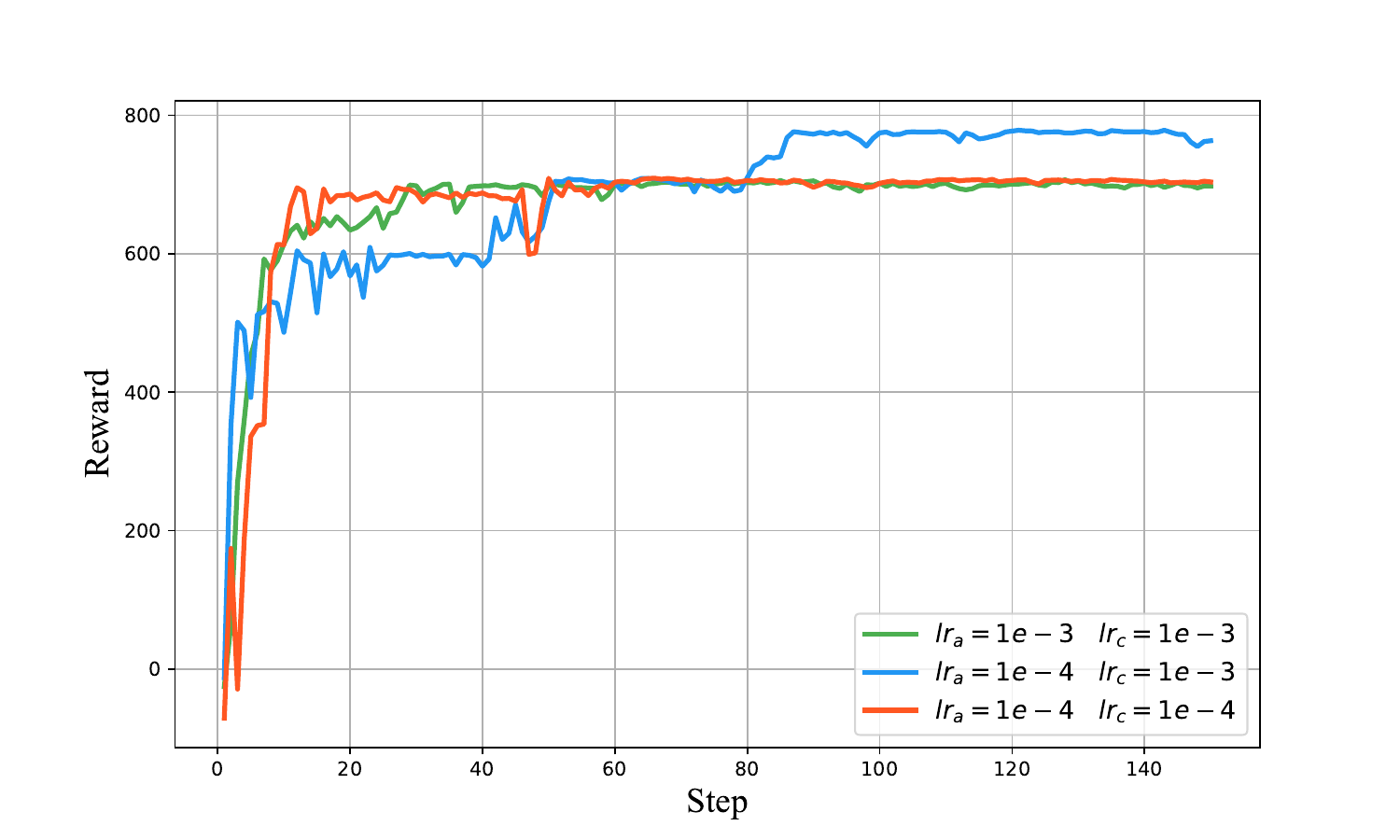}
    \caption{Convergence comparison of MUECRS Algorithm with different learning rates.}
    \label{different_learning_rate}
\end{figure}

Figure \ref{different_learning_rate} illustrates the convergence behavior of the MUECRS algorithm under different learning rates for the actor and critic networks. It can be observed that the algorithm converges in all three learning rate configurations. All three curves show a rapid increase during the initial 20 steps, with the curve corresponding to $lr_a=1e-4$ and $lr_c=1e-3$ exhibiting a slower rise compared to the other two curves. Between steps 60 and 80, all curves stabilize around a reward of 700. However, after approximately 80 steps, the curve for $lr_a=1e-4$ and $lr_c=1e-3$ begins to outperform the others, stabilizing at a higher reward of around 800. This indicates that, although the learning speed is slower initially with $lr_a=1e-4$ and $lr_c=1e-3$, it avoids premature convergence and ultimately achieves a higher cumulative reward. Therefore, in our experiments, we adopt the learning rates of $lr_a=1e-4$ for the actor network and $lr_c=1e-3$ for the critic network.

\begin{table}[ht!]
\centering
\caption{Major simulation parameters}
\label{Major_simulation_parameters}
\begin{tabular}{l|l}
\hline
\textbf{Parameter}  & \textbf{Value} \\ \hline
Number of user devices $M$  & 4 \\ \hline
Number of UAVs $N$  & 5$\sim$30 \\ \hline
Number of total time slots $T$ & 500 \\ \hline
The length of the area $X^{max}$ & 50m \\ \hline
The width of the area $Y^{max}$ & 50m \\ \hline
Minimum altitude of UAVs $Z^{min}$  & 10m \\ \hline
Maximum altitude of UAVs $Z^{max}$ & 20m \\ \hline
Minimum distance between UAVs $D^{min}$ & 3m \\ \hline
Data size of tasks $d_m$ & 100$\sim$150kb \\ \hline
CPU cycles per bit $c_m$ & 500$\sim$1000 cycles/bit \\ \hline
Maximum velocity of UAVs $v^{max}$ & 1.73m/s \\ \hline
Elevation angle of the UAV $\phi_n$ & $90^{\circ}$ \\ \hline
Environmental constants $a$ and $b$ & 9.61,0.16 \\ \hline
Bandwidth $B^{G2A}$  & 20MHz \\ \hline
Maximum transmit power of user $p_{m}$ & 1$\sim$1.2W \\ \hline
Maximum transmit power of UAV $p_{n}$ & 5W \\ \hline
Noise power $N_G$ & -70dBm \\ \hline
Computation resource of user $f_{m}$ & 0.8$\sim$1GHz \\ \hline
Computation resource of UAV $f_{n}$ & 10GHz \\ \hline
\end{tabular}
\end{table}

\begin{table}[ht!]
\centering
\caption{Major reinforce learning parameters}
\label{net_parameters}
\begin{tabular}{l|l}
\hline
\textbf{Parameter} & \textbf{Value} \\ \hline
Actor's hidden layer dimension $e_a$  & 64 \\ \hline
Critic's hidden layer dimension $e_c$ & 64 \\ \hline
Maximal training episode $episode_{max}$ & 250 \\ \hline
Learning rate for the actor network $lr_a$ & $10^{-4}$ \\ \hline
Learning rate for the critic network $lr_c$ & $10^{-3}$ \\ \hline
Soft update parameter $\tau$ & 0.01 \\ \hline
Discount factor $\gamma$ & 0.95 \\ \hline
Replay buffer size $B$  & $5\times 10^5$ \\ \hline
\end{tabular}
\end{table}

Table \ref{Major_simulation_parameters} presents the parameters related to the system model. We consider a 3D environment with a rectangular area of $50m \times 50m$ on the plane, where users are randomly dispersed. The minimum and maximum flight altitudes for the UAVs are $10m$ and $20m$, respectively. The environment contains $5 \sim 30$ users and 4 UAVs, with the initial positions of the UAVs set to $(0,0,10)$, $(0,50,10)$, $(50,0,10)$, and $(50,50,10)$. The UAVs have a coverage angle of $90^{\circ}$. Table \ref{net_parameters} displays the parameters related to reinforcement learning.

To demonstrate the effectiveness and robustness of the proposed algorithm, we compare the MUECRS algorithm with the following six algorithms:

\begin{itemize}
    \item CEJOMU algorithm \cite{10423388}: This algorithm uses the MADDPG algorithm, jointly deciding the 2D position movement of UAVs and the allocation of computing resources by considering three factors: UAV positions, UAV load balancing, and the distances between UAVs.
    \item Dueling DDQN (D3QN) algorithm \cite{10480601}: This algorithm uses the D3QN algorithm and discretizes the 3D flight of UAVs into six flight directions: $[a_1,a_2,a_3,a_4,a_5,a_6]$. Here, $a_1 = [0, 1, 0]$ represents forward flight, $a_2 = [0, -1, 0]$ represents backward flight, $a_3 = [1, 0, 0]$ represents leftward flight, $a_4 = [-1, 0, 0]$ represents rightward flight, $a_5 = [0, 0, 1]$ represents upward flight, and $a_6 = [0, 0, -1]$ represents downward flight.
    \item Random Trajectory (RT): This algorithm employs a random UAV trajectory, while the other offloading decisions and resource allocations are consistent with the proposed algorithm.
    \item All Offload (AO): This algorithm assumes that when users are within the UAV coverage area, all tasks are offloaded. The UAV trajectory and resource allocation are consistent with the proposed algorithm.
    \item All Local (AL): This algorithm assumes that all users adopt a local computation strategy.
\end{itemize}

\subsection{Simulation results and analysis}

\begin{figure}[!t]
    \centering
    \includegraphics[width=1\linewidth]{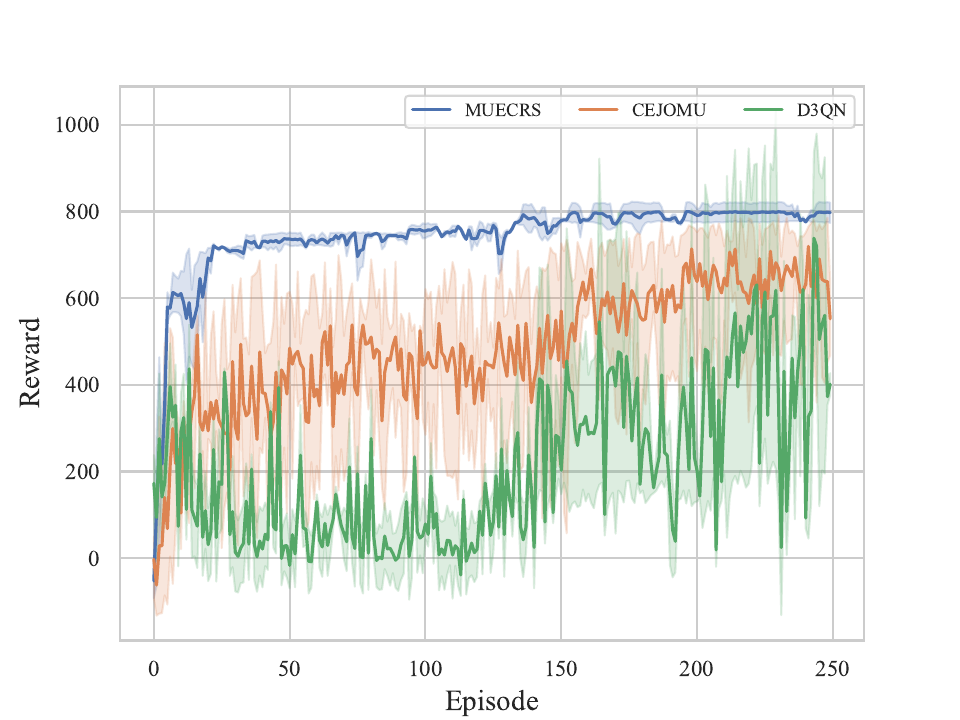}
    \caption{Convergence comparison of the three reinforcement learning algorithms.}
    \label{plot_smooth}
\end{figure}

Figure \ref{plot_smooth} shows the convergence comparison of the three reinforcement learning algorithms: MUECRS, CEJOMU, and D3QN, where the horizontal axis represents the number of training episodes, and the vertical axis represents the reward. As can be seen from the figure, the MUECRS algorithm performs the best in terms of convergence speed, stability, and optimization effect. The MUECRS algorithm rapidly increases the reward within the first 50 episodes, indicating a fast convergence speed, and stabilizes around 800 after 100 episodes. The CEJOMU algorithm shows a significant increase in reward within the first 50 episodes, but its convergence speed is slightly slower than that of MUECRS, stabilizing around 600 after 100 episodes. The D3QN algorithm shows some improvement within the first 150 episodes, gradually converging after 200 episodes, but with considerable fluctuations, stabilizing around 400. This indicates that the D3QN algorithm is more affected by the environment, and even after convergence, the reward still fluctuates significantly, suggesting that the strategy's performance varies greatly under different circumstances.

\begin{figure*}[!t]
    \centering
    \subfloat[\scriptsize MUECRS Algorithm]{%
        \includegraphics[width=0.32\linewidth]{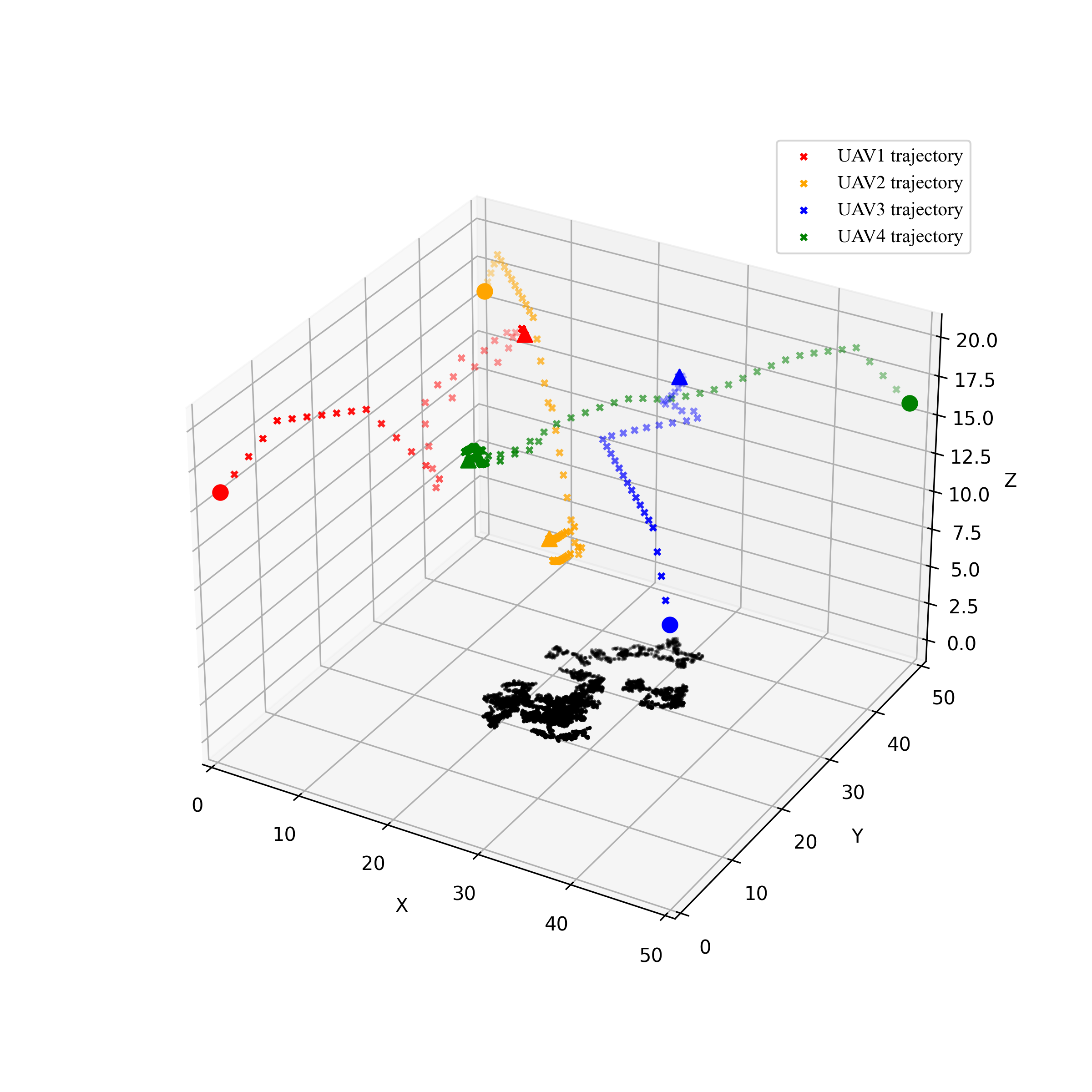}%
        \label{trajectory}
    }
    \hfill
    \subfloat[\scriptsize CEJOMU Algorithm]{%
        \includegraphics[width=0.32\linewidth]{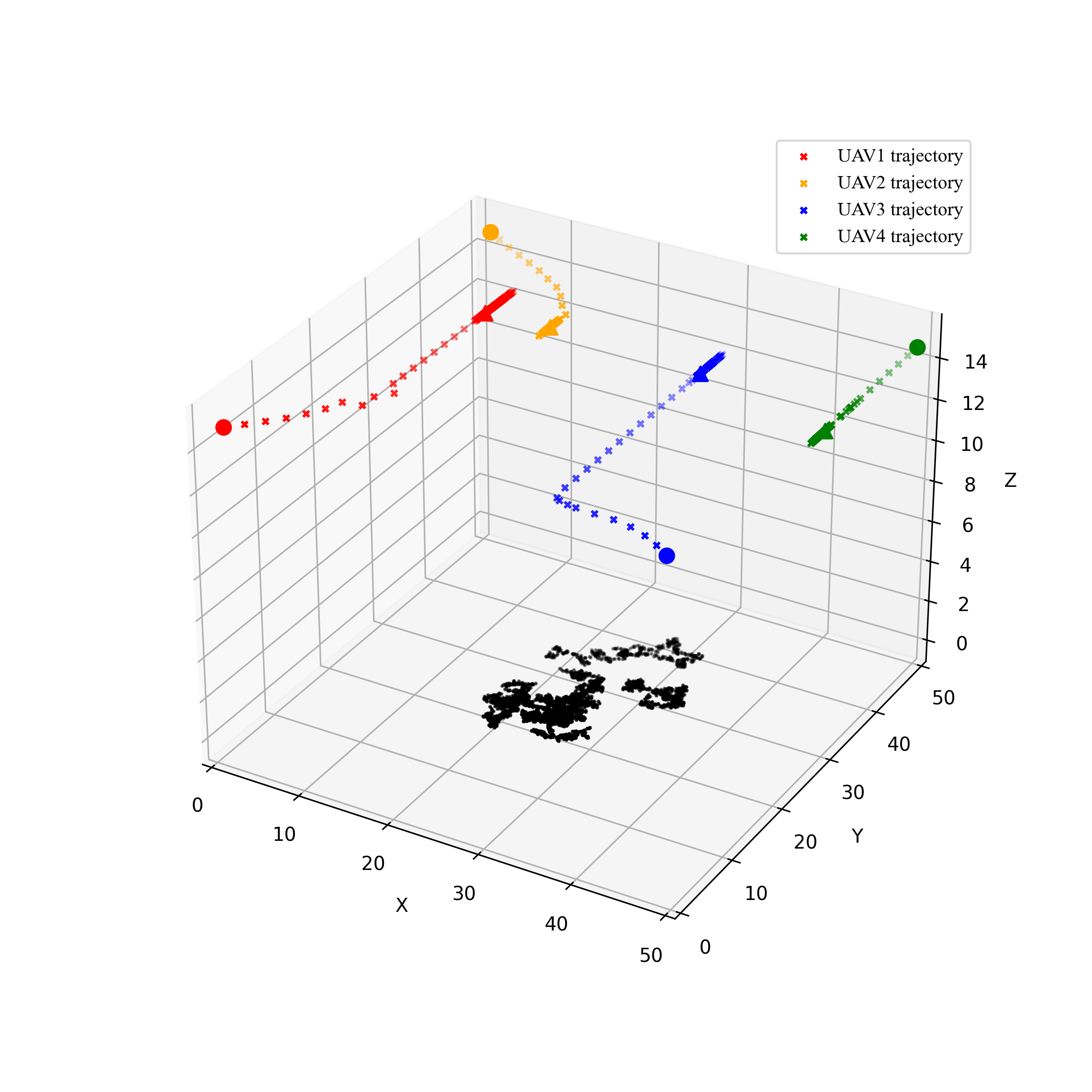}%
        \label{traj_CEJOMU}
    }
    \hfill
    \subfloat[\scriptsize D3QN Algorithm]{%
        \includegraphics[width=0.32\linewidth]{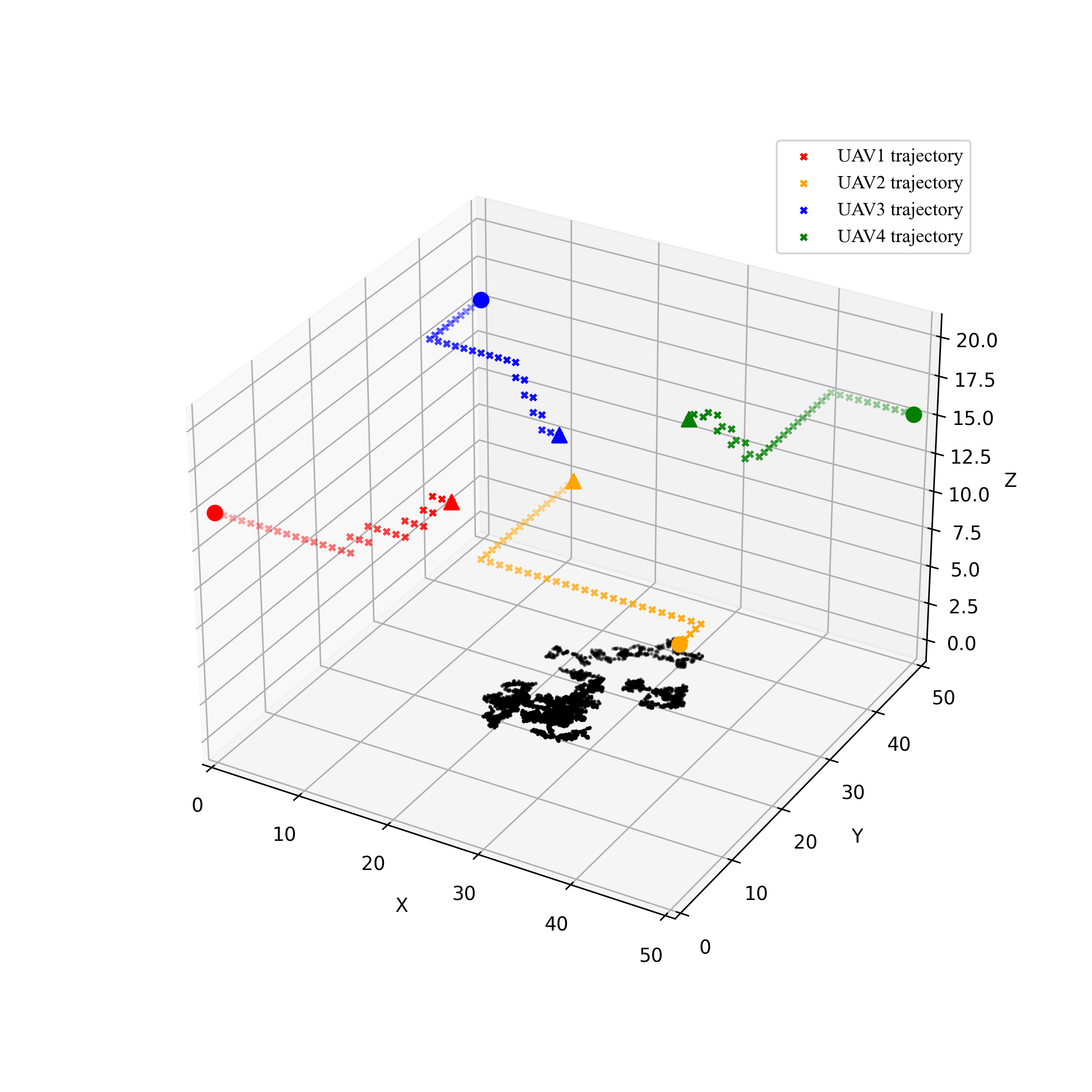}%
        \label{traj_D3QN}
    }
    \caption{Comparison of UAV flight trajectories using different methods.}
    \label{traj_comparison}
\end{figure*}

Figure \ref{traj_comparison} shows UAV flight trajectories of MUECRS, CEJOMU, and D3QN respectively. The flight area is a cubic region with dimensions $50m \times 50m \times 20m$. The red, yellow, blue, and green crosses represent the trajectories of the four UAVs, the circles represent the UAVs' starting points, the triangles represent the UAVs' endpoints, and the black dots represent the users' locations. Figure \ref{trajectory} shows the 3D flight trajectories of the UAVs using the MUECRS algorithm. It can be observed that all four UAVs fly from their starting positions towards the users and upwards, ensuring rapid coverage of the users. The increase in altitude allows the UAVs to expand their coverage area, thereby covering more users. After reaching the area above the users, the UAVs hover to provide stable coverage and enable quick response and task scheduling for the users. Figure \ref{traj_CEJOMU} shows the 2D flight trajectories of UAVs under the CEJOMU algorithm. Due to the dimensional limitations of the algorithm, the UAVs fly at the same altitude as their starting points. The algorithm also considers load balancing and the distances between UAVs, resulting in more dispersed flight trajectories that avoid overlap and resource wastage. However, this may lead to longer transmission delays during the user task transmission process. Figure \ref{traj_D3QN} shows the 3D flight trajectories of UAVs under the D3QN algorithm. The trajectories of each UAV exhibit 'stepped' changes, which are a result of the algorithm's discretization of flight directions. While this approach reduces the complexity of continuous path planning, it sacrifices some of the trajectory smoothness.

\begin{figure}[!t]
    \centering
    \includegraphics[width=1\linewidth]{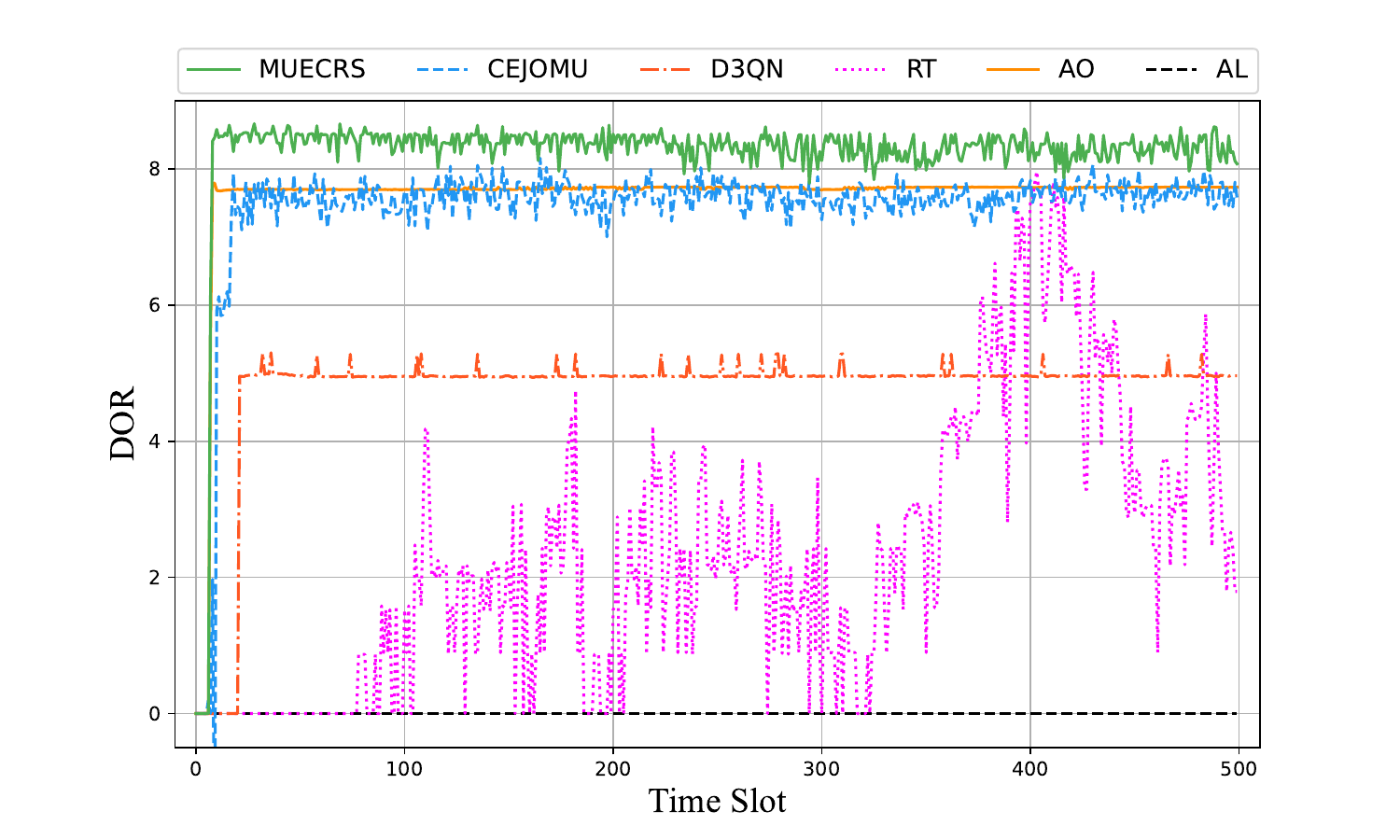}
    \caption{Comparison of DOR per time slot.}
    \label{plot_step_opt}
\end{figure}

Figure \ref{plot_step_opt} shows the comparison of DOR per time slot for different algorithms. The horizontal axis represents the time slots, with a total of 500 time slots in the experiment, and the vertical axis represents the DOR. It can be seen that the DOR for all algorithms is initially 0, which is due to the UAVs' starting positions not covering the users, preventing users from offloading tasks to the UAVs. As time progresses, the UAVs fly toward areas with a high concentration of users, achieving coverage and thus increasing the DOR. It can be observed that after training, the reinforcement learning algorithms achieve a relatively stable DOR after convergence. The MUECRS algorithm stabilizes the DOR around 8.5 by the fifth step, while the CEJOMU and D3QN algorithms stabilize around the tenth step, at approximately 7.5 and 5.1, respectively. This is because the CEJOMU algorithm uses a 2D flight mode, which is limited by flight altitude, and the D3QN algorithm's trajectory is based on a discrete action space, leading to suboptimal UAV trajectories and slower user coverage. The RT algorithm exhibits greater fluctuation due to the instability of UAV coverage caused by random trajectories.

\begin{figure}[!t]
    \centering
    \includegraphics[width=1\linewidth]{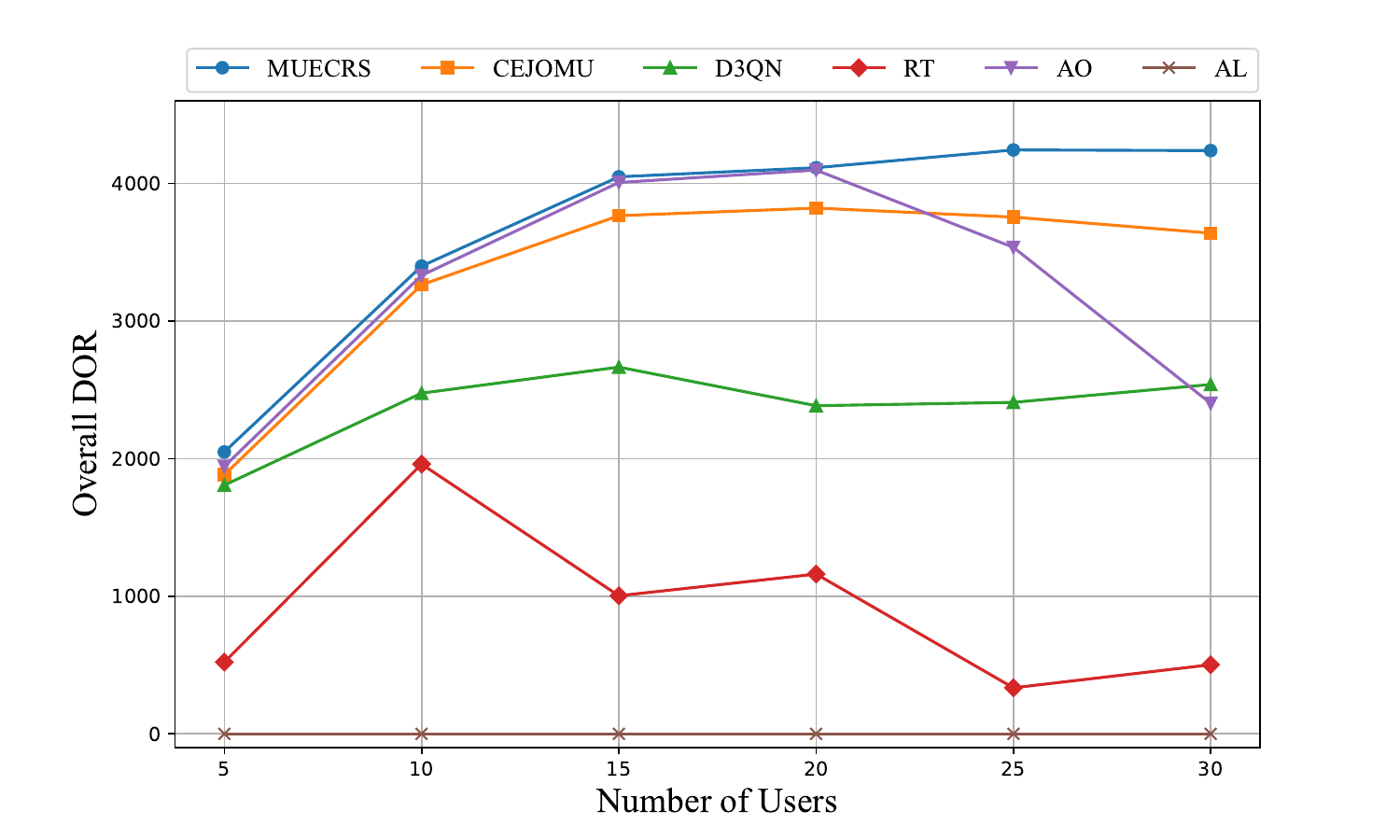}
    \caption{Comparison of overall DOR with different numbers of users.}
    \label{different_users}
\end{figure}

Figure \ref{different_users} compares the overall DOR for different numbers of users, with the horizontal axis representing the number of users, ranging from 5 to 30. It can be observed that as the number of users increases, the DOR of the MUECRS algorithm also increases and consistently outperforms the other algorithms. When the number of users reaches 30, the DOR of MUECRS is 4238.25, which is more than 16.7\% higher than the other algorithms and continues to show an upward trend. The CEJOMU and D3QN algorithms also see an increase in DOR as the number of users increases, but after the number of users exceeds 15, their DOR metrics stabilize around 3755.02 and 2408.71, respectively, with minimal improvement or even slight declines. The AO strategy experiences a significant decrease in DOR after the number of users exceeds 20 due to resource shortages and increased competition for MEC resources caused by excessive task offloading. The RT exhibits random fluctuations with changes in the number of users, while the AL strategy consistently maintains a DOR of 0.

\begin{figure}[!t]
    \centering
    \includegraphics[width=1\linewidth]{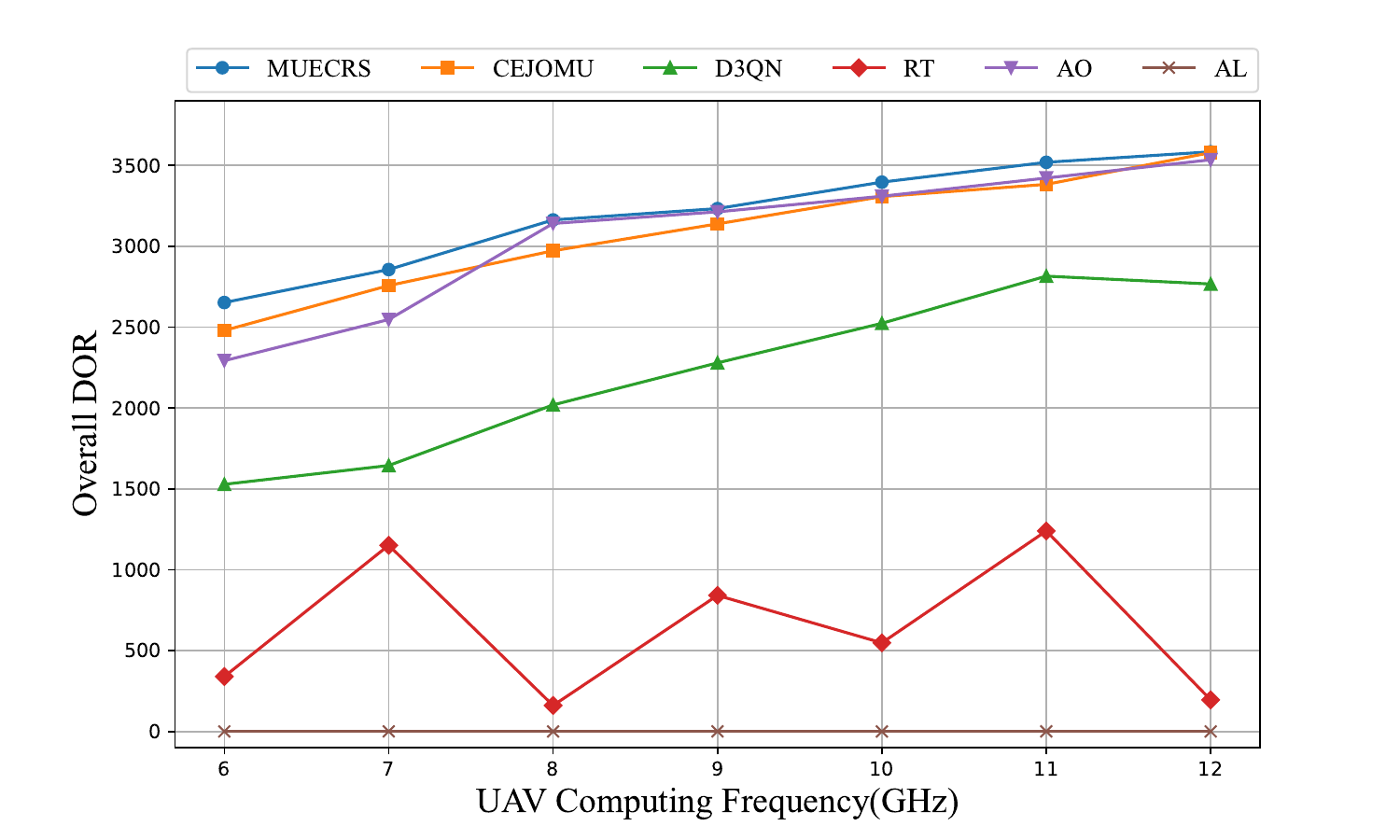}
    \caption{Comparison of overall DOR with different UAV computing capabilities.}
    \label{different_uav_f}
\end{figure}

Figure \ref{different_uav_f} shows the comparison of overall DOR with different UAV computing resources. The horizontal axis represents the UAV computing frequency, ranging from 6 GHz to 12 GHz, and the vertical axis represents the overall DOR. It can be observed that as MEC resources increase, the MUECRS and CEJOMU algorithms perform the best: the DOR for these two algorithms steadily increases with the rise in UAV frequency, consistently maintaining high values. Notably, the MUECRS algorithm exhibits the highest optimization values across the entire frequency range, indicating that as the UAV computing frequency increases, the processing speed of UAVs accelerates, enhancing the advantages of offloading and consequently increasing the DOR. The D3QN algorithm shows moderate performance. Although its optimization values also rise with frequency, the growth rate is relatively slower, and the overall DOR is lower than that of the MUECRS and CEJOMU algorithms. The RT algorithm performs the least consistently, with significant fluctuations in DOR across the frequency range, indicating poor stability of this strategy, and its overall optimization values remain relatively low.

\begin{figure}[!t]
    \centering
    \includegraphics[width=1\linewidth]{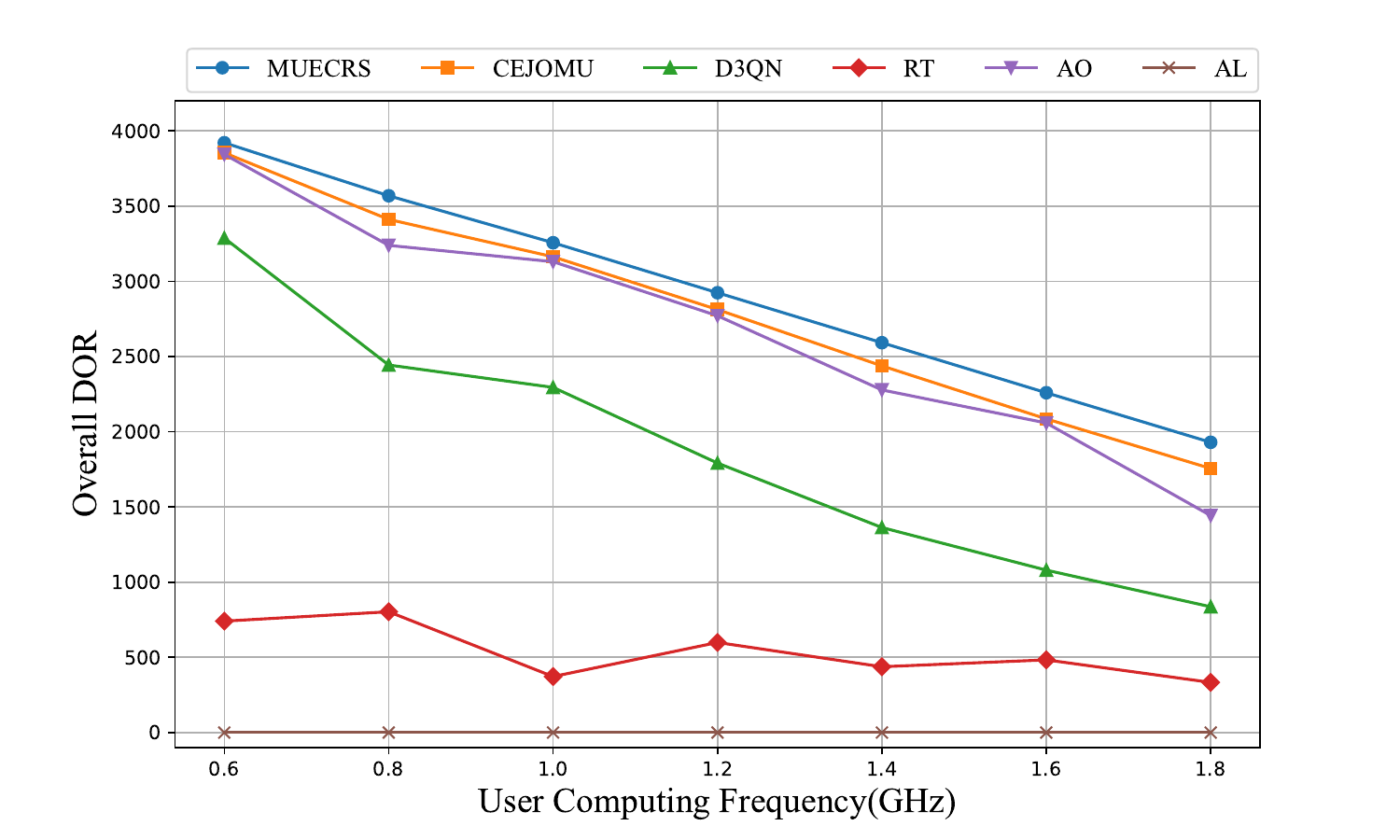}
    \caption{Comparison of overall DOR with different user computing capabilities.}
    \label{different_users_f}
\end{figure}

Figure \ref{different_users_f} shows the comparison of overall DOR with different user computing resources. The horizontal axis represents the user computing frequency, ranging from 0.6 GHz to 1.8 GHz, and the vertical axis represents the overall DOR. It can be observed that as user computing resources increase, local computing capability is enhanced, leading to a reduced need for offloading, and the DOR of all algorithms decreases accordingly. Among the algorithms, MUECRS performs the best, consistently exhibiting the highest DOR across the entire frequency range. The CEJOMU, D3QN, and AO algorithms also show high and stable DOR, but they are all lower than that of the MUECRS algorithm. The RT algorithm generally shows a downward trend with significant fluctuations, indicating poor stability. The analysis of the figure suggests that when user computing resources are limited, offloading tasks to MEC servers on UAVs can significantly reduce the computing burden on the user side, ensuring that tasks are completed quickly and efficiently.
 
\section{Conclusion}
In this paper, we have addressed the problem of maximizing overall DOR in a multi-UAV-assisted MEC scenario by designing the 3D flight trajectories of UAVs, making 0-1 offloading decisions for computing tasks, and scheduling channel bandwidth and computing resources. To solve this complex MINLP problem, we have proposed a low-complexity solution that combines the MADDPG algorithm, the CD algorithm, and KKT conditions. In the MADDPG algorithm, we have first modeled the problem as a POMDP, and then we have treated each UAV as an agent that decides its own flight trajectory. The CD algorithm and KKT conditions have been alternately applied, using a greedy strategy to search for resource allocation strategies. By utilizing KKT conditions, we have derived closed-form solutions for channel resource allocation and computing resource allocation. Finally, through extensive experimental comparisons with state-of-the-art algorithms, the proposed algorithm has outperformed comparative algorithms by at least $16.7\%$, demonstrating faster response times and better stability.

In our future work, we plan to extend our research by considering obstacles such as trees and buildings in real environments, which will make the modeling closer to real-world conditions and the research problems and algorithms more practical.


\bibliographystyle{IEEEtran}
\bibliography{ref}

\vfill
\end{document}